\documentclass[pre,aps,twocolumn,superscriptaddress,showpacs,floatfix]{revtex4-1}

\usepackage[english]{babel}

\usepackage{dcolumn}
\usepackage{amssymb}
\usepackage{amsmath}
\usepackage{bm}
\usepackage{color}
\usepackage{graphicx}
\usepackage[normalem]{ulem}

\usepackage[usenames,dvipsnames]{xcolor}

\usepackage{color}
\definecolor{darkblue}{rgb}{0,0,0.6}
\definecolor{darkred}{rgb}{0.6,0,0}

\definecolor{ao(english)}{rgb}{0.0, 0.5, 0.0}
 \newcommand{\so}[1]{}      
 \newcommand{\blue}[1]{#1}      
 \newcommand{\green}[1]{}       

\usepackage{hyperref}
\hypersetup{
colorlinks=true,			
urlcolor=darkblue,
citecolor=darkblue,		
linkcolor=MidnightBlue,	
}

\newcommand{\gdot}{{\dot{\gamma}}}
\newcommand{\xmin}{{x_{\tt min}}}
\newcommand{\br}{{\bm r}}

\begin{document}

\title{Properties of the density of shear transformations in driven amorphous solids}

\author{Ezequiel E. Ferrero} 
\affiliation{Instituto de Nanociencia y Nanotecnolog\'{\i}a, CNEA--CONICET, 
Centro At\'omico Bariloche, (R8402AGP) San Carlos de Bariloche, R\'{\i}o Negro, Argentina.}

\author{Eduardo A. Jagla} 
\affiliation{Centro At\'omico Bariloche, Instituto Balseiro, 
Comisi\'on Nacional de Energ\'ia At\'omica, CNEA, CONICET, UNCUYO,
Av. E. Bustillo 9500
R8402AGP S. C. de Bariloche, 
R\'io Negro, Argentina}

\date{\today}

\begin{abstract}

The strain load $\Delta\gamma$ that triggers consecutive avalanches
is a key observable in the slow deformation of amorphous solids.
Its temporally averaged value $\langle \Delta\gamma \rangle$ 
displays a non-trivial system-size dependence that constitutes one of the 
distinguishing features of the yielding transition.
Details of this dependence are not yet fully understood.
We address this problem by means of theoretical analysis and 
simulations of elastoplastic models for amorphous solids.
An accurate determination of the size dependence of $\langle \Delta\gamma \rangle$ 
leads to a precise evaluation of the steady-state distribution of local distances 
to instability $x$. 
We find that the usually assumed form $P(x)\sim x^\theta$ 
(with $\theta$ being the so-called pseudo-gap exponent) 
is not accurate at low $x$ and that in general $P(x)$ 
tends to a system-size-dependent \textit{finite} limit as $x\to 0$. 
We work out the consequences of this finite-size dependence standing 
on exact results for random-walks and disclosing an alternative 
interpretation of the mechanical noise felt by a reference site.
We test our predictions in two- and three-dimensional elastoplastic models,
showing the crucial influence of the saturation of $P(x)$ 
at small $x$ on the size dependence of $\langle \Delta\gamma \rangle$ and 
related scalings.
\end{abstract}

\maketitle

Punctuated dynamics is inherent to many out of equilibrium driven systems.
When energy is loaded at a small and fixed rate, the nature
of the system is such that this energy is dissipated in sudden bursts of activity 
typically called slip events or avalanches. 
This kind of systems are referred to as displaying a stick-slip dynamics. 
Examples include the relative motion of tectonic plates giving rise 
to earthquakes~\cite{carlson1994dynamics},
the sliding of charge density waves~\cite{FisherPR1998},
the driven movement of a magnetic interface 
in thin magnetic films~\cite{FerreCRP2013}, 
the intermittent motion of rain droplets on a windshield~\cite{DeGennesRMP1985} and 
the plastic rearrangements occurring in amorphous solids under a slow 
and sustained strain increase~\cite{NicolasRMP2018}. 
In all these cases, a stationary situation is established in which, on average, 
the stress (or energy) increase during quiescence periods is equal 
to the stress (or energy) drop released during avalanches.

Suppose that we drive a system with stick-slip dynamics on its steady state,
and we are interested in the statistics of strain increases needed to produce
a new slip event, for systems of different sizes. 
If the system consists on $N$ `blocks' that can be locally
destabilized, one expects that the load needed to trigger
the weakest block scales with $1/N$.
This is, if we double the system size, the closest instability
will be halfway apart in terms of strain increase needed.
Equivalently, if we drive the system at a small finite rate,
the pace at which we observe slip events doubles when we double
the system size.
More rigorously, if avalanches have a maximum extent that does not 
diverge as the system size goes to infinity, then the system is extensive.
The previously mentioned balance between accumulation and release of energy 
then implies that if the system size is doubled, the average load increase that has 
to be applied to generate a new avalanche is halved. 
While this is the case for most stick-slip phenomena
(e.g., friction, depinning, wetting, etc.), the
behavior of amorphous solids under deformation disobeys this logic.
In the deformation of amorphous materials, if we double the system size,
the rate at which we observe slip events does not double.
It increases, but less; it is \textit{sub-extensive} in the system size.
In other words, to trigger the next slip one needs to load
\textit{more than expected}.
As a consequence, when the system finally yields, the slip of a single
block is not enough to compensate the load excess and system spanning
avalanches of plastic events emerge.
Therefore, the plastic activity is rarely confined to localized 
plastic events and, instead, it is mostly originated in extended 
structures~\cite{MaloneyPRL2004} 
This points clearly to the non-extensiveness of the problem.
In fact, if we consider conversely that the dynamics of the problem 
produces system size spanning avalanches, then a doubling in the 
system size would not duplicate the number of avalanches.

It is now well established that the statistics of the mean strain load 
$\langle\Delta\gamma\rangle$ needed to trigger consecutive avalanches 
in the steady state of quasistatically driven amorphous solids has 
profound consequences on the criticality of the yielding 
transition~\cite{LernerPRE2009, Karmakar2010, KarmakarPRE2010b}.
In particular, its finite-size scaling is expected to be a manifestation
of the distribution of \blue{putative} shear transformation zones $P(x)$
\blue{($x \equiv \Sigma_{\tt th} - \Sigma$ stands for the local
`stress distance' to the local yielding threshold $\Sigma_{\tt th}$)} and bounds through 
scaling relations the possible exponents governing the 
avalanche size distribution~\cite{LinEPL2014, LinPNAS2014, budrikis2015universality, tyukodi2015depinning, liu2015driving, NicolasRMP2018}.
While consensus on this scaling being sub-extensive prevails, i.e., 
$\langle\Delta\gamma\rangle \sim N^{-\alpha}$, with $0<\alpha<1$,
there have been conflicting views on the value of $\alpha$ and its 
justification. 
On one hand molecular dynamics (MD) simulations~\cite{LernerPRE2009,Karmakar2010,KarmakarPRE2010b} 
of model glasses under quasistatic deformation support an universal 
value of $\alpha\simeq 2/3$, valid both in $d=2$ and $d=3$ dimensions. 
On the other hand, elasto-plastic (EP) models for amorphous 
solids~\cite{LinEPL2014,LinPNAS2014} display dimension-dependent values of $\alpha$.

\newpage

Despite large advances in the field, 
theoretical arguments have not taken yet in full 
account the statistical significance of the 
$\langle\Delta\gamma\rangle$ sub-extensivity
in the steady state.
Such a scaling has been well accounted as a justification 
for system-spanning avalanches of plastic activity in the 
system; but the fact that it also implies  an inherent 
discrete evolution for the `local distances to threshold' 
$x$ was disregarded. 
In this work, we address this issue, presenting an alternative and
consistent picture for the finite-size scaling of \so{$\langle\xmin\rangle$ (or  $\langle\Delta\gamma\rangle$)}
\blue{$\xmin \equiv \min_i x_i $.}
Standing on the ground provided by previous works~\cite{FerreroSM2019,FerreroPRL2019, FernandezAguirrePRE2018,jagla2018prandtl}, we interpret the evolution of the stress 
(and thus also $x\blue{_i}$) in a generic region of the system as an effective  
stochastically-driven random walk. 
Working out the finite-size scaling of the relevant discrete jump in this 
walk we derive a generic scaling law $\langle\xmin\rangle \sim N^{-\alpha}$
with $\alpha=2/3$.
In doing so, we revisit the significance and shape of the distribution $P(x)$,
which shows a finite limit $P_0 = \lim_{x\to 0} P(x)$ that scales as
$P_0 \sim N^{1-\alpha}$ in the thermodynamic limit.
This scaling occurs independently of the eventual value of $\theta$ 
observed at intermediate values of $x$ where $P(x)\sim x^\theta$ can 
be fitted.

In Sec.~\ref{sec:overview} we refresh the subject under discussion in a mini-review.
In Sec.~\ref{sec:randomwalks} we \so{we step back, onto}
\blue{motivate and perform an analysis in terms of} simple random-walkers problems 
with exact solutions 
to understand the effect of discrete steps.
In Sec.~\ref{sec:mechnoise} we rationalize the collective effect of plastic events 
during avalanches as an effective mechanical noise with a discrete step effect 
on the `walks' of local stresses.
In Sec.~\ref{sec:epmodels2Dand3D} we test our hypothesis in extensive simulations 
of 2D and 3D elastoplastic models, presenting rigorous finite size analysis for 
$\langle\xmin\rangle$ and $P(x)$ in different cases.
Finally, in Sec.~\ref{sec:conclusions} we summarize our results, that
we believe allow to construct a consistent scenario
for what seemed \textit{a priori} 
contrasting results in literature.

\begin{figure}[b]
\begin{center}
\includegraphics[width=0.99\columnwidth]{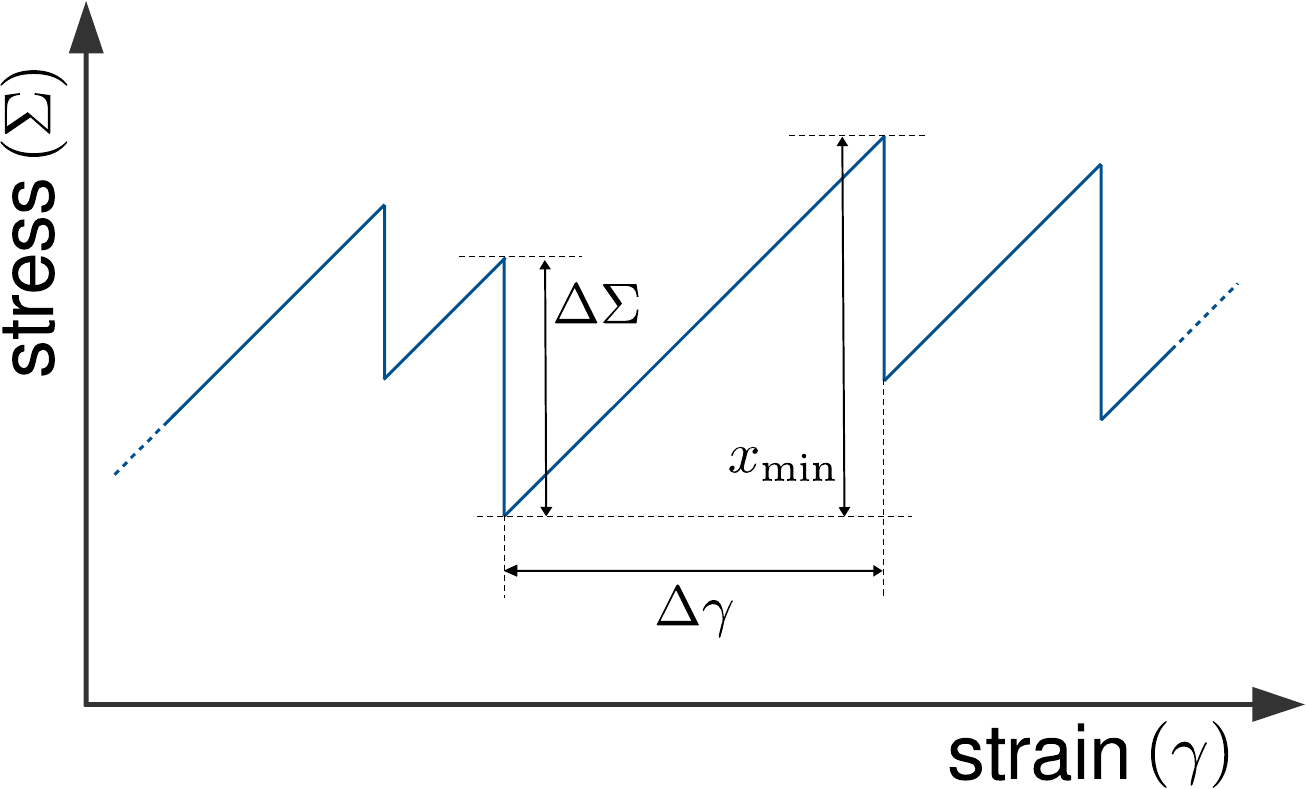} 
\end{center}
\caption{
Steady-state stress-strain scheme in the quasistatic shear deformation of an amorphous solid.
}
\label{fig:stick-slip}
\end{figure}

\section{Overview of the subject}
\label{sec:overview}

Let us start by briefly reviewing the main concepts and literature results 
on this topic.
\so{Consider the schematic stress-strain evolution}
\blue{In essence, one can think of a yielding material as a fully-connected 
set of elastoplastic blocks characterized by a local stress $\Sigma_i$ or, 
equivalently, a local {\it distance} to the stress threshold 
$x_i=\Sigma_{{\tt th}i} - \Sigma_i$}.
\blue{These blocks evolve according to a global load that drives the 
$x_i$ values towards zero. 
When a particular $x_j$ reaches zero, the block yields, reaching a new 
equilibrium position (at some new, positive value of $x_j$) while at 
the same time producing (via elastic interactions) a modification of 
the values of other $x_i$ all across the system. 
We say that these are `mechanical kicks' given to the blocks
each time one of the blocks yields. 
The yielding of block $j$ may produce (due to the mechanical kicks)
the yielding of other blocks in cascade. 
This is the origin of avalanches in the system that characterize the 
dynamics.
Because of this avalanche dominated dynamics, the stress-strain evolution 
in the system has a qualitative form as} depicted in Fig.\ref{fig:stick-slip}.
Once in its steady state, a driven amorphous solid performs an interspersed
sequence of load periods and slip events when the relevant stress component
is monitored.
A stationary average value is expected for the stress on a steady state. On top of this 
average value, stress fluctuations contain information on the physics of the problem.
In particular, the average strain increase of the loading periods (referred to as 
$\langle \Delta\gamma \rangle$) and the average stress-drop during the slip events 
($\langle \Delta\Sigma \rangle$) must be proportional in a stationary situation 
(see Fig.\ref{fig:stick-slip}), namely 
\begin{equation}
\langle \Delta\Sigma \rangle=B \langle \Delta\gamma \rangle
\label{sigma-gamma}
\end{equation}
with $B$ an elastic constant.

In a quasistatic athermal dynamics, \so{calling $x_i$ the value of additional
stress  that site $i$ needs to reach instability,} the stress increment 
that needs to be applied to trigger a new avalanche is nothing but the 
minimum $x_i$ across the system, \so{noted} \blue{namely} $\xmin$.
Then, 
\begin{equation}
\langle \xmin \rangle = \langle \Delta\Sigma \rangle .
\label{so}
\end{equation}
Further, energy drops, quantifying the energy dissipated during plastic
avalanches, can be easily related to the stress-drops as
$\langle \Delta U \rangle = \Sigma_Y \langle \Delta\gamma \rangle V = \frac{\Sigma_Y}{B} \langle \Delta\Sigma \rangle V$;
where $V = L^d =N$ is the system volume and $\Sigma_Y$ is the global yield stress~\cite{LernerPRE2009,Karmakar2010}.
In fact, the starting point of the current discussion can be traced back to a series 
of MD quasistatic simulations~\cite{LernerPRE2009,Karmakar2010,KarmakarPRE2010b}
where the following system-size scaling laws were verified
\begin{equation}\label{eq:UandDsigma_scaling}
\langle \Delta U \rangle \sim N^\delta \quad;\quad \langle \Delta\Sigma \rangle \sim N^{-\alpha}
\end{equation}
holding $\delta+\alpha=1$, with $\delta\simeq 1/3$ and $\alpha\simeq 2/3$, 
both in two and three spatial dimensions.
Also~\cite{Salerno2013} independently showed compatible results.
In Ref.~\cite{Karmakar2010}, when thinking on the distribution of possible
plastic events, an ansatz was introduced.
Arguing that the distribution of energy barriers felt in a quasistatic 
loading protocol should grow as a power-law, 
it was proposed for the distribution of local distances to instabilities
($x$ is the generic value of all $x_i$ across the system) the expression
\begin{equation}\label{eq:xtothetheta}
P(x)\sim x^\theta ,
\end{equation}
for small $x$ and with $\theta>0$.
Then the distribution of $\xmin$ 
\so {(the smallest $x_i$ across the system)} follows a Weibull distribution
\begin{equation}
P(\xmin)\sim \xmin^\theta\exp(-N\xmin^\theta),
\end{equation}
with its mean value scaling as
\begin{equation}\label{eq:xminfss}
\langle \xmin \rangle \sim N^{-1/(1+\theta)} 
\end{equation}
which provides a justification for the scaling of Eq.~\ref{eq:UandDsigma_scaling},
linking $\alpha$ and $\theta$:  
\begin{equation}\label{eq:alphathetarelation}
\alpha=\frac{1}{1+\theta}.
\end{equation}
Yet, notice that Eq.\ref{eq:xminfss} does not 
imply $P(x)\sim x^\theta$.
The small argument power-law form of the Weibull 
distribution for $P(\xmin)$ was verified in the statistics of
the `as-quenched' state or \textit{isotropic} solid state, both 
in $d=2$ and $d=3$ dimensions~\cite{LernerPRE2009,Karmakar2010,KarmakarPRE2010b}.
And it is in fact for this case that the ansatz (\ref{eq:xtothetheta}) was 
proposed~\cite{Karmakar2010}.
Nevertheless, these pioneer MD simulations couldn't easily access 
the whole distribution $P(x)$, and results where only presented 
for $P(\xmin)$ (or $P(\Delta\gamma)$).

Luckily, soon after the problem was addressed by EP model simulations 
measuring the full $P(x)$ distribution~\cite{LinEPL2014}.
In there, a plausible law $P(x)\sim x^\theta$ was found not only in 
the `as-quenched' state but also at the critical stress.
The $P(x)\sim x^\theta$ ansatz was subsequently extended to describe in 
EP simulations not only the steady state~\cite{LinPNAS2014}, 
but also the transient regime~\cite{lin2015criticality} where 
a statistics of extended avalanches was equally observed.
It was concluded that $\theta$, and therefore $\alpha$ according to 
the construction, should be dimension and system parameter 
dependent, which was formalized in an analytic mean-field 
approach~\cite{LinPRX2016}.
This theory has the virtue of formally catching a strain-dependence 
of $\theta$, a feature that is observed in the transient regime both in 
EP~\cite{lin2015criticality,LinPRX2016}
and MD~\cite{HentschelPRE2015,WenchengPRE2019,ShangPNAS2020,ruscher2019residual}
simulations.
In such transient, the values of $\theta$ observed are highly 
non-universal, depending on system preparation, system parameters
and dimension~\cite{LinPRX2016,LinPRE2018,WenchengPRE2019}.

In the construction summarized in~\cite{LinPRX2016}, $\alpha$ 
is expected to follow the same trend as $\theta$ all the way
from the `as-quenched' state to the steady-state, keeping
the relation $\alpha=1/(1+\theta)$, and binding $\alpha$
to be also highly non-universal.
Nevertheless, one naturally expects $\theta$ and $\alpha$ 
to stop depending on strain in the steady-state, and indeed the 
literature has collected from the beginning evidence for such 
expectation~\cite{Karmakar2010,LinEPL2014,LinPNAS2014}.
Moreover, we have recently showed that in that limit those
exponents are model-independent~\cite{FerreroSM2019} for a large
set of EP model rules; they do depend on dimension though.
So, at some point the variation of $\alpha$ and $\theta$ with 
strain should vanish.
How that happens, may be a matter of theoretical discussion itself.
For the time being, we will focus on the limit of large strains
where a self-consistent and stationary stick-slip phenomenon is 
expected to occur.

Interestingly, in contrast with the case of `as-quenched' systems,
the relation $\alpha=1/(1+\theta)$ does not seem to hold so well 
in the numerical results of the steady-state in EP models.
For example, in~\cite{LinPNAS2014} $\theta$ is reported to be
$\sim 0.57$ and $\sim 0.35$ respectively in $d=2$ and $d=3$,
while $\alpha$ results form the $\xmin$ scaling in
$\sim 0.67$ and $\sim 0.79$ for those 
cases~\footnote{
As a matter of fact, \emph{different} values for the exponent $\theta$
are presented in Ref.~\cite{LinPNAS2014} when either 
fitted from the $P(x)$ distribution or computed from the 
`extremal dynamics' (the scaling of $\left< x_\text{min} \right>$)
through Eq.\ref{eq:alphathetarelation}; 
``a difference presumably resulting from corrections to scaling'' 
according to the authors.
}.
More recent EP simulations~\cite{TyukodiPRE2019} show
$\alpha \simeq 0.675$ combined with $\theta \simeq 2/3$ in $d=2$. 
And in~\cite{FerreroSM2019} we have observed $\alpha\simeq 2/3$ 
and $\theta\simeq 0.75$ for 6 different $d=2$ EP models, pushing
the relation $\alpha=1/(1+\theta)$ even further away from validity.
The apparent violation of such relation in the steady state
is accompanied by two related observations.
First, it is well known from the beginning of this discussions
that $P(\xmin)\sim (\xmin)^\theta$ does not show up in the 
steady-state~\cite{Karmakar2010,LinEPL2014};
in fact that law, valid for the `as-quenched' state,
is rapidly suppressed as soon as the applied stress is 
finite~\cite{HentschelPRE2015,WenchengPRE2019}.
Secondly, recent numerical results in both MD simulations~\cite{ruscher2019residual}
and EP models~\cite{TyukodiPRE2019,FerreroSM2019} have consistently 
made evident that in the steady state $P(x)$ displays a \so{prominent}
plateau at small values of $x$ \blue{(a non-zero base value, unmistakable 
in a double logarithmic plot $P(x)$ vs $x$)}, and suggested that the finite-size 
scaling of $\langle\xmin\rangle$ can be dominated by the behavior with system 
size of the finite asymptotic value of $P(x)$ at vanishing $x$ rather than
by \blue{the exponent} $\theta$.

What seems to be clear, at least, is that the plain assumption 
$P(x)\sim x^\theta$, which was somehow inherited from the `as-quenched' 
phenomenology and carried by for all values of strain, 
is insufficient in the steady state.
Nevertheless, for instance, a scaling relation based on Eq.~\ref{eq:alphathetarelation},
and linking the exponents $\tau, d_f$ that describe the distribution of 
avalanche sizes with the exponent $\theta$ (namely,
$\tau=2-\frac{\theta}{1+\theta}\frac{d}{d_f}$) 
has been largely adopted in the the EP models 
literature~\cite{liu2015driving,budrikis2015universality,karimi2016inertia,NicolasRMP2018},
always accepted without further justification and relying sometimes
on generous error bars for the exponents.
Something is missing in the understanding of what controls $\alpha$,
which might cause that even the latter relation among exponents should
be revised.
In this work we address the issue, admittedly limiting ourselves 
to the steady-state case, were we expect universal values of 
$\alpha$~\cite{Karmakar2010,HentschelPRE2015,FerreroSM2019}.

\subsection*{A mean-field approach to yielding}
\vspace{-0.3cm}

In \cite{LinPRX2016}, Lin\&Wyart \blue{extending a work
by Lemaitre and Caroli~\cite{lemaitre_arxiv2007}}
presented a mean-field approach  
\so{that explicitly considered for the firsts time the long-
tail nature of the mechanical `kicks' produced by plastic
events.}
\blue{which is based in the assumption that the mechanical 
kicks produced by yielding sites 
on every other site can be taken from a given distribution 
defined once and for all, independently of the state of the 
system, and, more importantly, that this distribution is heavy-tailed.}
\blue{
The mean-field dynamics can be described as follows.
If at time $t$ the block $j$ yields (reaches $x_j\leq 0$),
it is re-injected at a positive value of $x_j$ (e.g. $x_j=1$) and the rest 
of the blocks suffer mechanical kicks $\xi_i$ taken form a distribution $w(\xi_i)$
(see Eq.~\ref{eq:mechnoise_wyart}) with zero mean $\langle \xi_i \rangle = 0$,
\begin{eqnarray}
    x_j(t+1) & = & 1, \nonumber \\
    x_i(t+1) & = & x_i(t) + \xi_i - \frac{1-x_j(t)}{N-1}, \label{eq:rw_eom}
\end{eqnarray}
\noindent where the last term grants stress conservation and in this case 
the re-injection point has been chosen to be $x=1$.
}

\begin{figure}[t!]
\begin{center}
\includegraphics[width=1\columnwidth]{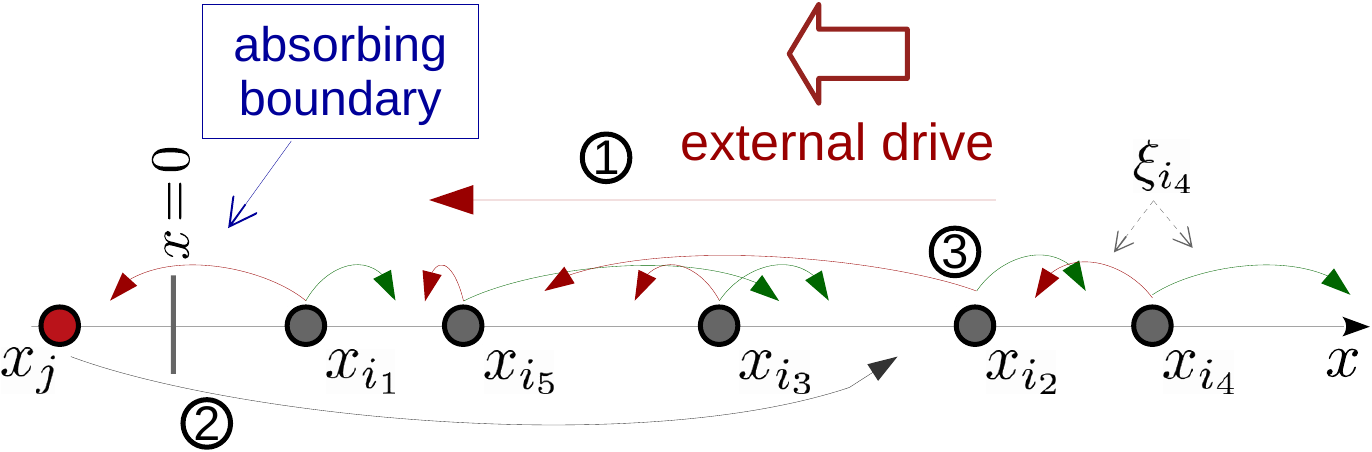} 
\end{center}
\caption{
\blue{Schematic wandering of the $x_i$ values.
The external drive pushes everyone towards zero {\large\textcircled{\small 1}}.
When $x_j$ yields, it is re-injected {\large\textcircled{\small 2}}
at a positive value of $x$ and all the remaining $x_i$ 
receive kicks $\xi_i$ {\large\textcircled{\small 3}}. 
These kicks can be either positive or negative
and of different intensities (as illustrated
by the arrows' different colors and lengths), 
according to the prescribed distribution $w(\xi)$.
The signed nature of the mechanical noise
allows the system to `forecast' the boundary~\cite{LinPNAS2014} 
and creates a density depletion of $P(x)$
close to $x=0$.
}
}
\label{fig:rw_for_x}
\end{figure}
\blue{
A highly distinctive feature of yielding phenomena lays in the fact 
that the mechanical noise distribution $w(\xi_i)$ comprises both 
positive and negative values.
This feature has its roots in the Eshelby response observed upon plastic 
events in amorphous materials and described in App.~\ref{sec:models}.
Once this is guaranteed, these mean-field models behave qualitatively 
like elastoplastic models of amorphous solids~\cite{NicolasRMP2018}: 
There is a global yield stress $\Sigma_Y$ such that for 
$\Sigma<\Sigma_Y$ the dynamics eventually stops, corresponding 
to the solid phase.
For, $\Sigma>\Sigma_Y$ the dynamics does not stop and is characterized
by a global strain rate.
}
\blue{The dynamics therefore can be rationalized as `random walks' 
of the elastoplastic blocks in the $x$-coordinate, 
with an absorbing boundary at $x=0$~\cite{LinPRX2016}, 
as qualitatively depicted in Fig.~\ref{fig:rw_for_x}.
This representation of the yielding phenomenon allows us 
to start by analyzing simple random-walk processes and 
extrapolate conclusions from there, see Sec.~\ref{sec:randomwalks}.
}

\blue{
The signed nature of the mechanical noise gives rise to a density
depletion of $P(x)$ close to the absorbing boundary~\cite{LinPNAS2014,FerreroSM2019}.
In contrast to the case of a purely positive interaction among sites
where each destabilized block tend to destabilize the others, the signed
kicks allow some blocks to escape the boundary and survive longer without 
yielding.
This is the hand-waving argument for the existence of a pseudo-gap
$P(x)\sim x^\theta$ with $\theta>0$.
}
\so{They studied the existence of a non trivial $\theta$ exponent 
as a consequence of the mechanical noise that different parts 
of the system produce on a particular site.}
\so{They}\blue{Formally solving the stochastic problem
of Eq.~\ref{eq:rw_eom}, Lin\&Wyart} 
concluded that $\theta$ depends continuously on the applied shear 
stress, non monotonically and without signs of universality at the yield 
stress\so{; what}\blue{. The latter observation} \so{left no hope}
\blue{leaves little room} for the \so{comparison of} \blue{expectation of universal} 
exponents among different EP models, not to talk about MD simulations\blue{. We 
will contrast this view}.

\so{Their construction~\cite{LinPRX2016} was based on the assumption of a 
mechanical noise $w(\xi)$ 
of the form
$w(\xi) \sim  \frac{1}{|\xi|^{\mu+1}}$,
in the particular case of $\mu\!=\!1$, that they claimed is the 
only value with physical meaning to be expected to occur. 
}

\subsection*{Alternative views}

\blue{The mean field construction in~\cite{LinPRX2016} is based on the assumption of a 
mechanical noise $w(\xi)$ 
of the form
\begin{equation}\label{eq:mechnoise_wyart}
w(\xi) \sim  \frac{1}{|\xi|^{\mu+1}}
\end{equation}
in the particular case of $\mu\!=\!1$, that the authors claim is the 
only value with `physical meaning' to be expected to occur. 
}

The necessity of the particular value $\mu=1$ has been recently questioned
\cite{FernandezAguirrePRE2018,FerreroSM2019,FerreroPRL2019}. 
In particular, 
it was argued the assumption $\mu=1$ in Eq.\ref{eq:mechnoise_wyart} is not in 
agreement with the observation of sub-extensive avalanches dominating the plastic 
activity in the quasistatic limit~\cite{FerreroSM2019,FerreroPRL2019}.
Other values of $\mu$ with $1<\mu<2$ acquire physical meaning 
after the mechanical noise is properly redefined (see Sec.\ref{sec:mechnoise}).
And, in fact, a value of $\mu \simeq 3/2$ was found to be consistent with the 
mechanical noise sensed numerically in six different EP models in two 
dimensions~\cite{FerreroSM2019}.
\blue{Recently, the physical case of $1<\mu<2$ has been also 
addressed in~\cite{parley2020aging} both for the aging and 
steady regimes, finding no reasons to discard it.}
Interestingly, the mean field theory of~\cite{LinPRX2016} 
yields a well defined value of $\theta=\mu/2$ when $1<\mu<2$,
independent on other parameters.
Yet, a uniquely-defined value of $\theta=\mu/2$ 
would still fail to explain through Eq.~\ref{eq:alphathetarelation} 
the value of $\alpha$ observed in both MD and EP simulations.
Remarkably, it has been recently observed quite clearly 
in both EP~\cite{TyukodiPRE2019,FerreroSM2019} and MD~\cite{ruscher2019residual}
simulations that the true shape of $P(x)$ at small values of $x$ 
deviates from a pure power-law $\sim x^\theta$, and has a finite 
limit
\begin{equation}\label{eq:P0plusxtheta}
P_0\equiv \lim_{x\to 0}P(x)\ne 0 .
\end{equation}
Namely, for any finite system size, $P(x)$ has a \textit{plateau} 
(when $P(x)$ is presented in a logarithmic plot) at small enough $x$.
As we will argue in the following,
the plateau in $P(x)$ is originated in the discrete nature of the 
mechanical noise that produces the ``kicks'' felt by $x_i$. 
These kicks (that we consider to be generated by extended plastic 
avalanches elsewhere in the system) push each $x_i$ to perform a 
(non-Gaussian) random walk.
In this scenario, it is the system size scaling of $P_0$ what
dominates the scaling of $\langle \xmin \rangle$ and 
controls the values of $\alpha$ and $\delta$ in 
Eq.~\ref{eq:UandDsigma_scaling}, which now turn to be 
compatible with the independent exponent $\theta=\mu/2$.

In the following, we elaborate on this picture. 
Our work deals largely with providing analytical arguments and numerical support
for the system size dependence of $P_0$ and $\langle \xmin \rangle$
and conciliates them with the existence of a well defined value
of $\theta$ that indeed describes an intermediate region of $x$ values
where $P(x) \sim x^\theta$.

\section{Simple random walks and the $P(x)$ plateau}
\label{sec:randomwalks}

\begin{figure}[t!]
\begin{center}
\includegraphics[width=0.97\columnwidth]{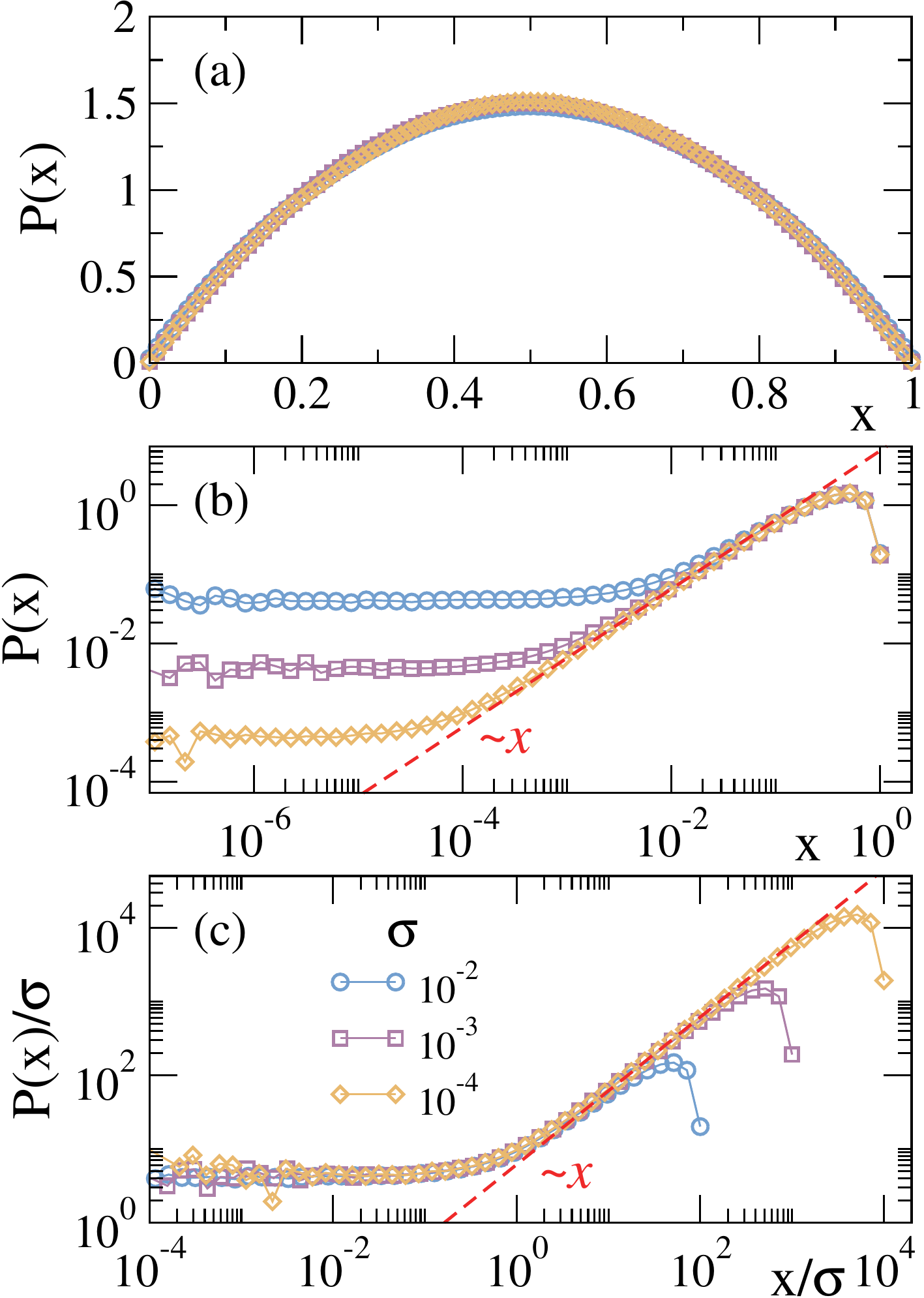} 
\end{center}
\caption{
Numerically determined probability distribution for a variable $x$ performing a standard (Gaussian) RW
in the interval $(0,1)$ with absorbing boundary conditions and random reinjection.
Different curves correspond to different values of the width of the single step distribution,
as indicated.
(a) Linear scale.
(b) Log scale to emphasize the behavior at low $x$.
The straight line shows the expected asymptotic limit for $\sigma\to 0$.
In (c) the axis are rescaled with $\sigma$ to show that the value $P(0)$
scales as $\sigma$.
}
\label{fig:RWgaussian}
\end{figure}

We analyze first a simple case. 
Consider a variable $x_i$ performing a \so{standard} random walk 
in the interval $[0,1]$, with absorbing boundary conditions.
When $x_i$ moves out of the interval, it is ``absorbed"
and re-injected in some random way~\footnote{In the simulations presented here the
reinjection is made randomly and uniformly in the full interval $(0,1)$}.
In the case of a continuous time random walk \blue{(a Wiener process)}, and when the reinjection is done
proportionally to the local value of the probability, 
the form of the distribution of
$x_i$ values observed along time in the steady state can be analytically 
computed to be $P(x)=\frac\pi2 \sin (\pi x)$.
For small $x$ it behaves as $P(x)\sim x$, i.e. $\theta=1$.
If we consider $N$ variables ($N \gg 1$) performing the same random walk, 
the minimum among them will be in the region in
which $P(x)$ is linear, and we will have 
$P(\xmin) \sim \xmin \exp(-N \xmin)$, 
and $\langle \xmin\rangle \sim N^{-1/2}$.

\begin{figure}[t!]
\begin{center}
\includegraphics[width=0.98\columnwidth]{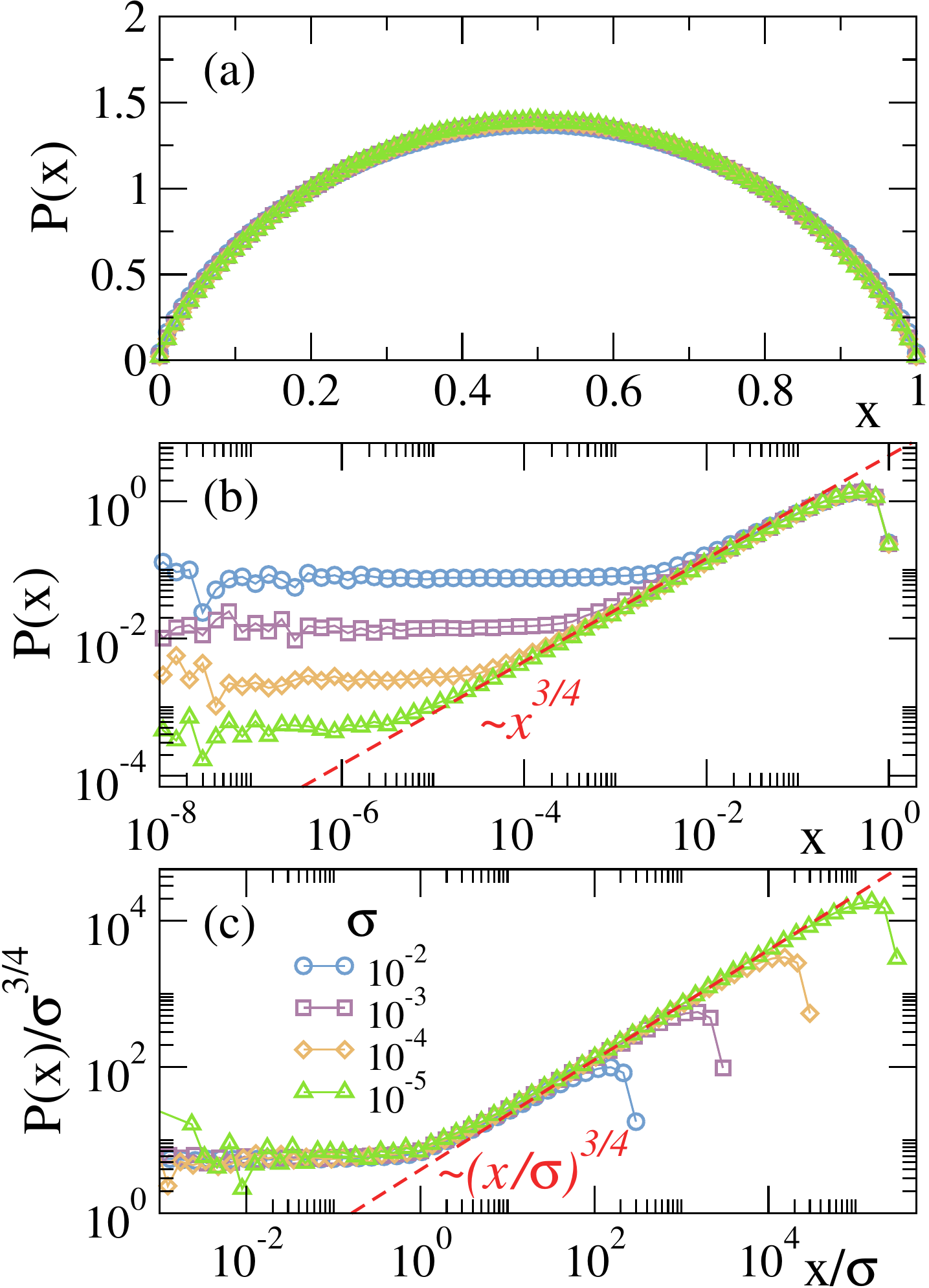} 
\end{center}
\caption{
Probability distribution for a variable $x$ performing a RW with Hurst exponent
$H=2/3$ in the interval $(0,1)$ with absorbing boundary conditions and random re-injection.
Different curves correspond to different values of the width of the single step distribution.
(a) Linear scale.
(b) Log scale to emphasize the behavior at low $x$.
The straight line shows the expected asymptotic limit for $\sigma\to 0$.
In (c) the axis are rescaled with $\sigma$ to show that the value $P_0$
scales as $\sigma^{-3/4}$ ($\sigma^{-1/2H}$ for a generic $H$).
}
\label{fig:RWheavytails}
\end{figure}

\so{However,} 
\blue{The random walks that we introduced in the previous section to 
describe the phenomenological dynamics of yielding are inherently discrete, 
and one needs to investigate the consequences of this fact on $P(x)$.
In fact, for}\so{For} a {\it finite step} random walk, \so{the situation changes.
A}\blue{a}lthough the overall form of $P(x)$ is the same as before, 
there is a small correction at small $x$ that depends on the step size, 
and has a strong effect on the value of $\xmin$.
Let's think for a moment of a particle performing a \emph{discrete} random-walk 
characterized by a step that is Gaussian-distributed, with a dispersion $\sigma$.
Assuming the particle is at some position $x_0$ at a given step, the next jump
makes $x$ to be distributed as $P(x)\sim\Theta(x)\Theta(1-x) \exp[-(x-x_0)^2/2\sigma^2]$,
where the Heaviside functions $\Theta$ appear because of the absorbing boundary conditions.
We note that the value of $P(0^+)$ is finite.
It turns out that this effect remains in the full solution for the stationary form of $P(x)$.
So, the discrete nature of the steps  taken by $x_i$
suffices to explain the finite limit of $P(x)$ as $x\to 0$.
In Fig.\ref{fig:RWgaussian} we see the distribution of $P(x)$ for Gaussian random
walks with different magnitudes of the average elementary step, namely,
different width $\sigma$ of the Gaussian `kicks'.
Fig.\ref{fig:RWgaussian}(a) shows the stationary distributions in lin-lin scale,
Fig.\ref{fig:RWgaussian}(b) shows them in log-log scale, and the scaling proposed in
Fig.\ref{fig:RWgaussian}(c) shows that the value of $P_0$ is proportional to $\sigma$.

The situation is conceptually identical in the case in which we consider
generalized RWs with a non-trivial Hurst exponent $H$;
this is, random walks generated by jumps $\xi$ drawn from a
heavy-tails distribution of the form
\begin{equation}
w(\xi) \sim \frac 1{|\xi|^{\frac 1H + 1}}, 
\label{wxi}
\end{equation}
for large $|\xi|$ with $1/2<H<1$.
Note first of all that in this case, the `typical jump' or distribution width 
$\sigma$ cannot be defined as being variance of the distribution because of 
its heavy tails, but it can be alternatively defined as 
$\sigma\equiv \langle{|\xi|}\rangle$.
As it was the case for a Gaussian variable, \so{in the  continuous time limit
(or vanishing small jumps, i.e., $\sigma\to 0$) RWs,} 
\blue{in the limit of vanishingly small jumps (i.e., $\sigma\to 0$) RWs,}
the form of
$P(x)$ for $x$ close to zero is still expected to be $P(x)\sim x^\theta$, 
where now $\theta=1/(2H)$~\cite{LinPRX2016}.
Yet, for finite $\sigma$, a finite value for $P_0$ appears, as shown in 
Fig.\ref{fig:RWheavytails} for $H=2/3$.
For concreteness, in this numerical example 
we have taken the distribution $w(\xi)$ to be given by 
Eq.\ref{wxi} if $|\xi|>\xi_0$, and $w(\xi)=0$ if $|\xi|<\xi_0$. 
This distribution has a width $\sigma=\langle{|\xi|}\rangle= \xi_0/(1-H)$.
We see that the limiting value of $P_0$ as a function of $\sigma$ 
scales as $P_0\sim \sigma ^{3/4}$ (Fig.\ref{fig:RWheavytails}c) . 
In the generic case with $1/2<H<1$, $P_0$ scales as
\begin{equation}
 P_0\sim \sigma^\theta.
\label{us2h}
\end{equation} 
This can be justified by noticing that close to $x=0$, $\sigma$ is the 
only possible scaling quantity with the same dimension as $x$.
Then we can write
\begin{equation}
P(x) = P_0 f(x/\sigma)
\end{equation} 
with $f(u) \simeq  1$ for $u \ll 1$.
On the other hand, 
for $x\gg \sigma$ (but still `small') we must have $P(x)\simeq C x^\sigma$ with $C$ 
independent of $\sigma$, therefore implying Eq.~\ref{us2h}. 
In other words, $\sigma$ marks a scale crossover below which the distribution
of $x$ values tends to a constant~\footnote{We thank D. Vandembroucq for 
pointing this out.}.

Finally, notice that everything we have said for the steady state
distribution $P(x)$ populated along time is also true if we populate
the distribution with the $x_i$ values of many independent walks
in their steady state.

\subsection{$N$ random walks without or with drift}
\label{sec:NrandomWalkersWdrift}

Let's consider then $N$ independent random walkers subject to
the following protocol.
Now, starting form a condition where every $x_i$ is in the interval $(0,2)$
we look for the minimum $x_i$, that we indicate as $\xmin$.
Every site is shifted by an amount $-\xmin$.
The site resulting with $x_i=0$ is re-injected in the box at $x_i=1$ and 
everyone updated by a (randomly) signed random quantity $\xi$ taken from 
a distribution similar to Eq.~\ref{wxi} ($\mu=1/H$)
\begin{equation}\label{eq:mechnoise_wyart_mod}
w(\xi) = \frac{A}{M_N} \frac{1}{|\xi|^{\mu+1}},
\end{equation}
but with upper and lower cutoffs
set to $\xi_{\tt up} = (2A/\mu)^\frac{1}{\mu}$ and 
$\xi_{\tt lo} = (2A/\mu)^\frac{1}{\mu} M_N^{-\frac{1}{\mu}}$ for 
it to be normalized~\cite{LinPRX2016}.
Importantly, here $M_N$ is an $N$-dependent parameter, frequently 
chosen as $N$ itself (see~\cite{LinPRX2016}). 
In the simulations of this toy model we will use $M_N=1/\xmin$ for reasons that will 
be clearer later on~\footnote{Note 
that instead of using at each step the latest $\xmin$ to define the distribution 
$w(\xi)$, one could use the self-tuned mean value $\left<\xmin\right>$ and 
the conclusions are identical.}.
Every site resulting in $x_i \leq 0$ (and eventually in $x_i \geq 2$) 
after the random kicks is also re-injected at $x_i=1$ 
(but \textit{not} producing further kicks).
The $N$ walkers feel these kicks independently, yet they are drifted globally 
by $-\xmin$ after each kick update.
In order to clearly identify the effect of such global drift, we will also 
analyze the case where we avoid the global drift step and simply: re-inject
the site with the minimum $x_i$, give random kicks to everyone and 
further re-inject those that go out of the box.

\begin{figure}[t!]
\begin{center}
\includegraphics[width=0.518\columnwidth]{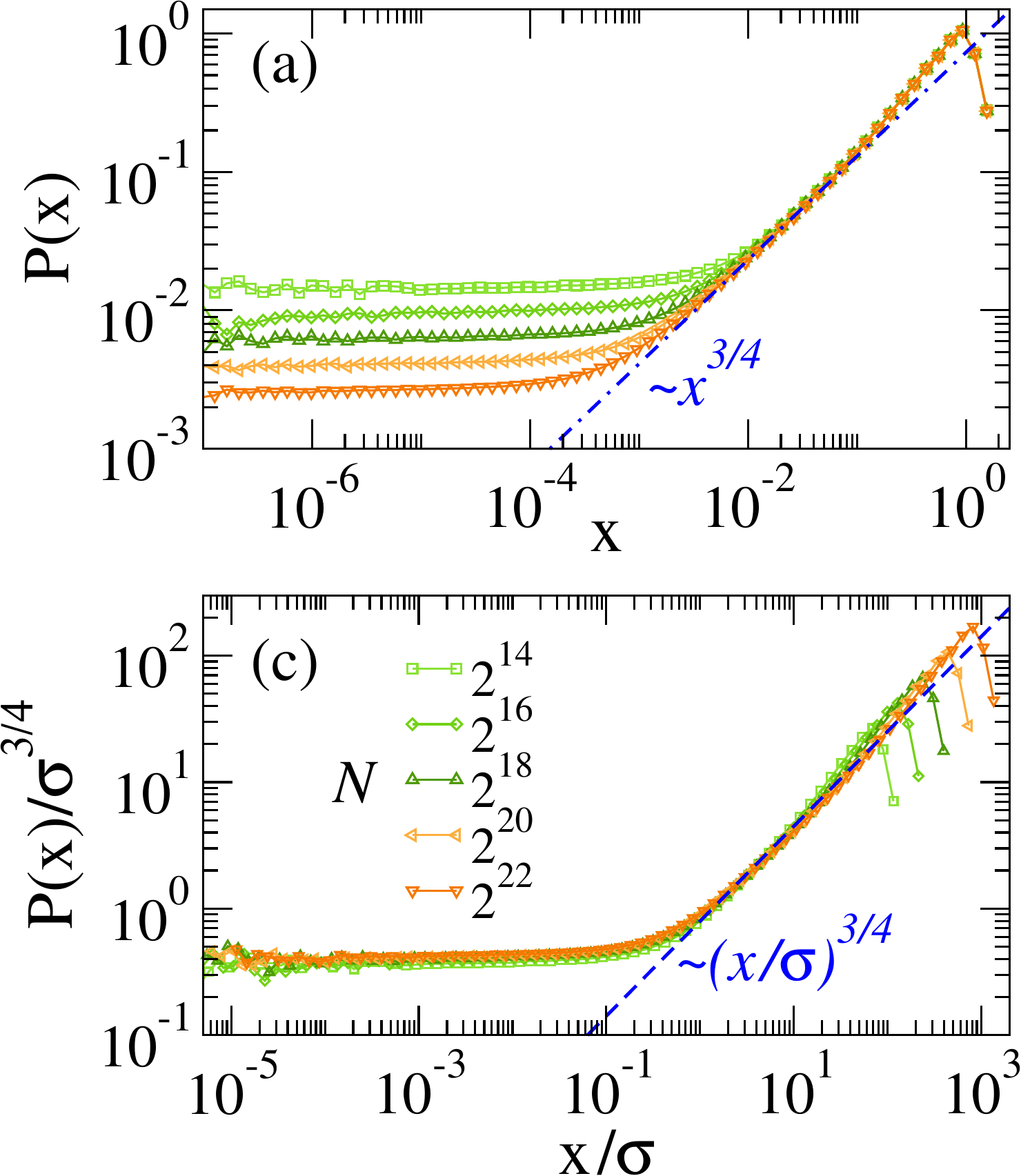}
\includegraphics[width=0.469\columnwidth]{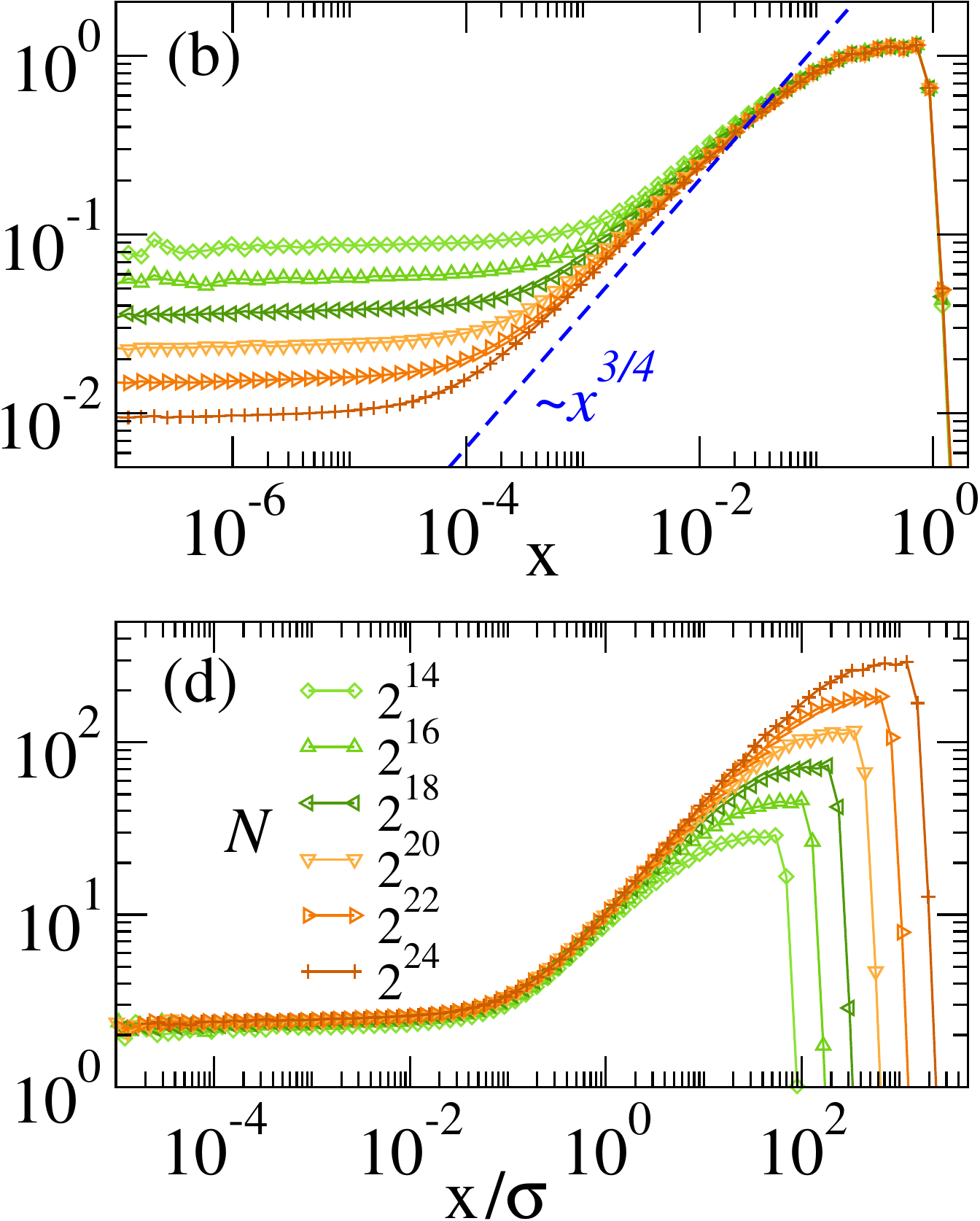} 
\end{center}
\caption{
Probability distribution populated by $N$ variables $x_i$ subject
to a dynamics of random kicks taken from a heavy tailed distribution
like the one in Eq.~\ref{eq:mechnoise_wyart_mod},
with Hurst exponent $H=2/3$ and $A=0.1$. 
Sites absorbed at the boundaries are re-injected in $x=1$.
Different curves correspond to different $\sigma \sim M_N^{-1/\mu} \sim N^{-4/9}$
{\it Left:} Without drift: (a) raw-data, (c) rescaled with $\sigma$ to show $P_0\sim \sigma^{-1/2H}$. 
{\it Right:} With drift. (b) raw-data, (d) rescaled with $\sigma$.
}
\label{fig:plkicks_A0.1}
\end{figure}

In both protocols, with and without drift, a steady state is established after 
a transient and the resulting $P(x)$ distributions are shown in Fig.\ref{fig:plkicks_A0.1}.
We can see that the drift couples the dynamics of the walkers 
and produces the effect of a `belly' on the curves that delays 
the decrease of $P(x)$ as we sense $x$ decreasing. 
The choice of the parameter $A$ now becomes relevant. 
If $A$ is small, the drift effect overtakes good part of the $P(x)$
distribution and it masks the power-law regime which gets difficult 
to determine, forcing us to simulate very large systems (or very small $\sigma$). 
If instead $A$ is big enough (closer to 1) the drift effect
is much diminished (data not shown).
In any case, when a reasonably large power-law region is
granted, the $\theta$ exponent is preserved for any $A$, 
$\theta=\mu/2$ provided that $1<\mu<2$. 
Notice that, despite this `belly' effect, the existence of a plateau at small
$x$ is unchanged, and the predictions $P_0\sim \sigma^\theta$ still holds,
as can be seen in the data collapse of Fig.~\ref{fig:plkicks_A0.1}(b,d).

The dynamics that we have just described can be though as a 
mean-field model for a system of elasto-plastic blocks with 
local thresholds where each of them feels an external drive
and a noise represented in $w(\xi)$.
We will now analyze a spatially extended system of driven interacting
blocks in this context.

\section{Effective mechanical noise of an interacting system and
the $P(x)$ distribution}
\label{sec:mechnoise}

Let us imagine a coarse-grained representation of an amorphous
material under deformation, represented by a scalar stress $\Sigma_i$
on each block and local yielding thresholds $\Sigma_{{\tt th}i}$.
The variables of interest will be the local distances to threshold 
$x_i = \Sigma_{{\tt th}i} -\Sigma_i$.
Our argumentation line is based on the analysis of the mechanical noise 
felt by a given site of such a system, caused by the plastic activity
elsewhere and governing the `wandering' of $x_i$.

For the results of the previous section to be applicable to the
present case, this noise must consist \textit{ideally} of independent, 
uncorrelated kicks.
As previously mentioned, Refs.~\cite{LinPRX2016,LinPRE2018} 
present a mean-field model considering kicks of a mechanical noise
generated by single Eshelby events. 
We will refer to these kicks generated by single sites as `elementary' kicks. 
The approximation of Ref.~\cite{LinPRX2016,LinPRE2018} describes qualitatively well 
the overall phenomenology observed in numerical simulations, 
but fails in predicting the exponents observed, at least below $d=4$.
This discrepancy was indeed ascribed to the presence of ``dimensional effects''
or correlations between the elementary kicks produced in different positions of 
the system.
We believe that the quantitative predictive power of this kind of analysis can 
be improved, still keeping the ``mean-field'' character of the approach, by 
noticing and taking into account that elementary kicks are not independent. 
Elementary kicks produced by sites that participate of the same avalanche are highly 
correlated among them, but those from different avalanches are not. 
This fact allows us to build a mean-field approach based on independent 
non-elementary kicks.
One possible choice is to define them as the integrated kicks given by 
avalanches, that in the quasistatic limit are by definition uncorrelated events.

The fact that the uncorrelated mechanical noise under consideration is 
produced by {\em avalanches} as a whole is the reason why now
$\mu$ in Eq.~\ref{eq:mechnoise_wyart_mod} can be different 
from the value $\mu=1$ that was obtained considering the effect of 
hypothetical uncorrelated elementary kicks instead~\cite{LinPRX2016}.
Actually, this alternative approach of avalanche-level noise
was already followed in~\cite{FernandezAguirrePRE2018,FerreroSM2019}.
Simulations of different EP models in two dimensions produce in a test site a 
noise characterized by a Hurst exponent $H\simeq 2/3$; which from the point of view
of the mechanical noise is equivalent to consider that such noise is taken randomly
from a distribution like Eq.~\ref{eq:mechnoise_wyart_mod} with $\mu \simeq 3/2$.
With that being proved to be effectively the case for a fully 
interacting system~\cite{FernandezAguirrePRE2018,FerreroSM2019}, 
we cannot expect anything different for its full distribution $P(x)$ 
than the features discussed in previous sections. 

\subsection*{Finite size scaling of the $P(x)$ plateau}

The  mechanical noise represented by Eq.~\ref{eq:mechnoise_wyart_mod}
contains as a fundamental parameter the value of $\mu$ (or $H\equiv 1/\mu$).
A second property of the distribution that has an important physical 
impact is its ``width'' $\sigma$. 
In particular, we are interested in how it scales with system size $N$.
The lower cutoff of the distribution 
$\xi_{\tt lo}\!=\!(2A/\mu)^\frac{1}{\mu}\!M_N^{-\frac{1}{\mu}}$ is
related to the system size and fixed by normalization. 
If $1\!<\!\mu\!<\!2$, the width $\sigma$ can be shown to be
proportional to $\xi_{\tt lo}$, and so
\begin{equation}
\sigma \sim M_N^{-1/\mu}
\label{sigma_M}
\end{equation}
It will be fact the finite-size behavior of the lower cutoff in 
$w(\xi)$, the noise produced by the far away plastic activity,
what will dominate the scaling of interest.
There is also an upper cutoff for the kick distribution, $\xi_{\tt up}$, 
but that is related with the strongest, nearest plastic events,
and independent on the system size~\cite{LinPRX2016,LinPRE2018}.

We have shown in the previous section that any finite step random walk process 
of a variable $x$ with absorbing boundaries, subject to such a random noise with 
$1\!<\!\mu\!<\!2$ implies that in the steady state 
\begin{eqnarray}
P(x)&\sim& x^\theta~~~~\mbox {for} ~~~ x\gtrsim \sigma \label{ppp0}\\
P(x)&\sim& \sigma^\theta~~~~\mbox {for} ~~~ x\to 0
\label{ppp}
\end{eqnarray}
where $\theta=\mu/2$, and $\sigma$ is the ``width'' of the 
distribution $w(\xi)$, as previously defined.
For instance, possible functional forms for $P(x)$
at small $x$ are
$P(x) \simeq \sigma^\theta + x^\theta$ or 
$P(x) \simeq (\sigma + x)^\theta$.
Furthermore, we have shown that $N$ random walkers, 
coupled by a common global drift 
generate the same limiting form of $P(x)$ as $x\to 0$. 

\begin{figure}[t!]
\begin{center}
\includegraphics[width=1.0\columnwidth]{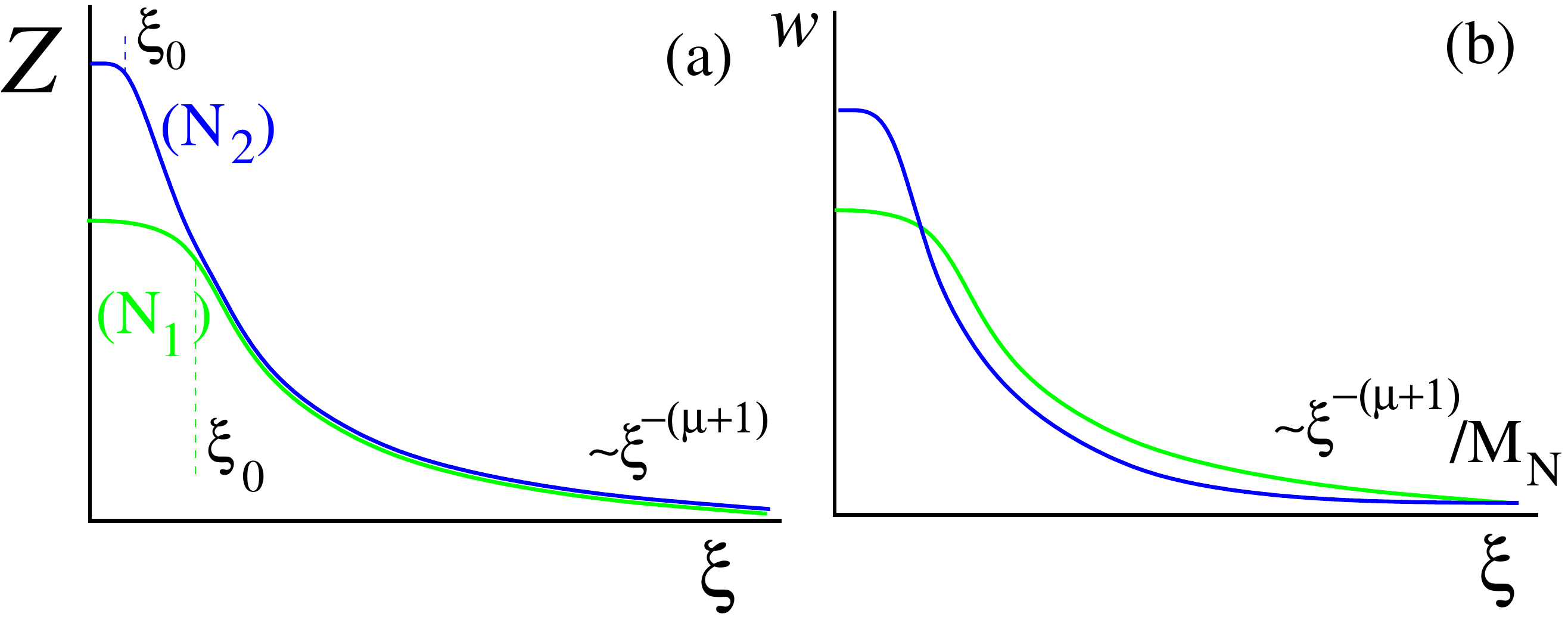} 
\end{center}
\caption{
(a) A schematic plot of the number density of kicks $Z$ of a given 
intensity $\xi$ observed in systems of two different sizes $N_1$, and $N_2>N_1$.
The two curves differ below the small size threshold $\xi_0$ but are coincident 
in the heavy tail part, for large $\xi$.
(b) The two curves in (a) normalized to become the probability distribution $w(\xi)$.
The normalizing factor is the number $M_N$ of avalanches that occur in the two systems
under the same increase of external strain.
}
\label{fig:sketch}
\end{figure}

The missing ingredient to make connection with the actual mechanical noise
felt by a given block in an amorphous solid 
is to work out the explicit dependence of $M_N$ in Eq.~\ref{eq:mechnoise_wyart_mod}
~on the system size $N$, and use it to calculate the scaling of
$\sigma$ (Eq.~\ref{sigma_M}) and thus the $N$-dependence of 
$\langle \xmin \rangle$.
Note that the approach of~\cite{LinPRX2016,LinPRE2018} uses $M_N = N$ which 
implicitly considers that each of the $N$ sites produces independent kicks 
on the generic block $i$, perturbing $x_i$.
We would like to stress here that this is clearly not realistic.
Furthermore, in careful consideration, it goes itself against the 
basic feature of yielding phenomena displaying size-spanning 
avalanches and sub-extensive scaling for the rate of plastic events.
Using $M_N=N$ and $\mu=1$ in Eq.\ref{eq:mechnoise_wyart_mod} 
implies somehow extensivity if kicks are supposed to be independent.
Instead, we think on the total noise produced by one avalanche.
Among the marginal kicks that a site receives (the ones that it almost fail 
to catch because of working in a finite system $N$), the dominant one is not 
the kick coming from a single site at the maximum possible distance, 
but the largest possible kick coming from such a distance.
That is, a kick coming from the largest avalanche at the largest distance.
If Eq.~\ref{eq:mechnoise_wyart_mod}
~represents the distribution of kicks generated by 
individual avalanches in the system, 
the value of $M_N$ must be chosen in accordance with this interpretation.

The dependence of $M_N$ on system size $N$ can be worked out as follows.
Consider two systems with different sizes $N_1$ and $N_2>N_1$, and suppose 
that we want to compare the number of kicks of intensity $\xi$ produced
onto some reference site when a fixed (long) deformation strain is applied 
to the system.
The Eshelby interacting kernel decays in space as $\sim 1/r^d$, and this 
implies that increasing the system size from $N_1$ to $N_2>N_1$ does not 
produce new large kicks~\footnote{In fact the largest kicks are produced by 
neighbor avalanches, the coarse-grained lattice description imposes the upper 
cutoff of the kick distribution, the minimal distance.}, 
but instead increases the number of small ones, those generated at large
distances in the system with $N_2$ sites. 
This means that if we plot the density number of kicks observed at a given 
site as a function of the kick magnitude, we would obtain a plot as 
the one qualitatively depicted in Fig.~\ref{fig:sketch}(a).
The portion of these curves following the $1/|\xi|^{\mu+1}$ law
will be mostly indistinguishable for the two system sizes.
Now, in order to plot the probability distribution $w(\xi)$, 
as shown in Fig.~\ref{fig:sketch}(b), it is clear that we have 
to divide by the total number of avalanches (kicks) that occurred 
in each case.
This is why $M_N$ in Eq.~\ref{eq:mechnoise_wyart_mod}
~must be considered to be proportional to such a number.
In other words, $M_N$ and the average size of avalanches in the system, 
noted $\overline S$, must be related through
\begin{equation}
M_N\sim N {\overline S}^{-1} \sim \langle \Delta\Sigma\rangle ^{-1}
\label{eq:mdsigma}
\end{equation}
(which, together with Eq.~\ref{so} justifies our choice for $M_N \simeq 1/\xmin$ 
in the toy model of the previous section \ref{sec:NrandomWalkersWdrift}). 
Now, collecting the results of Eqs. \ref{so}, \ref{sigma_M}, and \ref{eq:mdsigma}
we arrive at the important result\footnote{See Appendix B for the discussion of
a case in which the assumptions made to derive this result do not apply, 
and then Eq.~\ref{eq:sigma_xmin} does not hold.}

\begin{equation}\label{eq:sigma_xmin}
\sigma  \sim \langle \xmin \rangle^{1/\mu}.
\end{equation}
Introducing this into Eq.~\ref{ppp} we get
\begin{equation}
P_0 \sim \sigma^\theta \sim  \langle \xmin \rangle^{1/2},  
\label{eq:P0_xmin}
\end{equation}
since, for $1<\mu<2$, $\theta /\mu=1/2$~\cite{LinPRX2016}.
Remarkably, this result is independent of $\mu$ in such range.

We are now only one step away from our general 
scaling results.
As mentioned before, recent results in simulations of different 
EP models~\cite{TyukodiPRE2019,FerreroSM2019} and also
in MD simulations~\cite{ruscher2019residual} have shown that 
(i) a plateau exists for $P(x)$ at vanishing $x$, 
but also that (ii) $\langle \xmin \rangle$ shifts 
towards the plateau region of $P(x)$ as the 
system size $N$ is increased.
This can now be analytically justified:
From \ref{ppp0} and \ref{ppp} 
the crossover between the plateau and the power-law region 
is expected at $x_{\tt cross}\simeq \sigma$. 
Combined with Eq.~\ref{eq:sigma_xmin}, 
this provides $x_{\tt cross} \sim \langle \xmin \rangle^{1/\mu}$.
For any $\mu>1$, this tells that $\langle \xmin \rangle$
becomes lower than $x_{\tt cross}$ for large $N$.
In practice, crossovers can be very broad,
yet, in the limit $N\to \infty$ the following relation holds
\begin{equation}\label{eq:oneoverN}
\langle \xmin \rangle P_0 \simeq 1/N
\end{equation}
Using Eqs.~\ref{eq:P0_xmin} and \ref{eq:oneoverN}
we finally obtain the two important predictions:
\begin{equation}\label{eq:xmin_scaling}
\langle \xmin \rangle \sim N^{-2/3}
\end{equation}
and
\begin{equation}\label{eq:P0_scaling}
P_0 \sim N^{-1/3}.
\end{equation}

\noindent Notice further that if we assume $P(x) \simeq P_0 + x^\theta$, 
using Eq.~\ref{eq:P0_xmin}: 
$P(\langle \xmin \rangle) \simeq \langle \xmin \rangle^{1/2} + \langle \xmin \rangle^\theta$.
And, provided $\theta=\mu/2>1/2$, the second term becomes negligible over 
the first when $\langle\xmin\rangle$ is small enough.
We then could also expect a good ansatz to be:
\begin{equation}\label{eq:Pxmin_xmin}
P(\langle \xmin \rangle) \simeq \langle \xmin \rangle^{1/2}.
\end{equation}
Followed up from Eq.~\ref{eq:P0_xmin}, the latter would interchange
$P_0$ by $P(\langle \xmin \rangle)$ in every subsequent expression.
In the limit $N\to\infty$ both formulations are equivalent, since we expect
$P(\langle \xmin \rangle)$ to be part of the plateau and identical to $P_0$.
Notice nevertheless that Eq.\ref{eq:Pxmin_xmin} (and the ones 
derived from it) may work well even before reaching that limit.

The scaling provided by Eqs. \ref{eq:xmin_scaling} and \ref{eq:P0_scaling}
(or alternatively $P(\langle \xmin \rangle)\sim N^{-1/3}$) 
is quite generic, as it does not depend on the actual value 
of $\mu$ neither on the dimension of the problem. 
Even more, it is highly stimulating, since it agrees with the original 
observations of the $\langle \xmin \rangle$ scaling in MD simulations~\cite{LernerPRE2009,Karmakar2010} both in $d=2$ and $d=3$.
Yet, there are assumptions implicitly made in their deduction 
that can limit their validity.
For instance, our construction does not account for anisotropy
effects on the dimensions composing the system, which could
affect the scaling of any observable with the global system 
size $N$. 
Such an effect appears clearly when considering three dimensional 
systems, as we discuss bellow.
In addition, Eqs. \ref{eq:xmin_scaling} and \ref{eq:P0_scaling} do 
not apply in the case of a model with a (quenched) random kernel,
that we describe in Appendix~\ref{sec:random_k}, mainly due 
to the failure of the argument about the scaling of $\sigma$ with $N$.
In the next section we test the predictions of Eqs.\ref{eq:xmin_scaling}
and \ref{eq:P0_scaling} in 
elasto-plastic models in dimensions $d=2$ and $d=3$.

\section{Elasto-plastic models in 2 and 3 dimensions}
\label{sec:epmodels2Dand3D}

We now present results of quasistatic simulations of 
spatially extended elasto-plastic models.
We will limit ourselves in particular to the 
Picard's model~\cite{Picard2005}.
Details about model definition and simulation protocols can 
be found 
in the Appendix ~\ref{sec:models}, and data was
produced with essentially the same codes used in~\cite{FerreroSM2019}.

\begin{figure}[t!]
\begin{center}
\includegraphics[width=1\columnwidth]{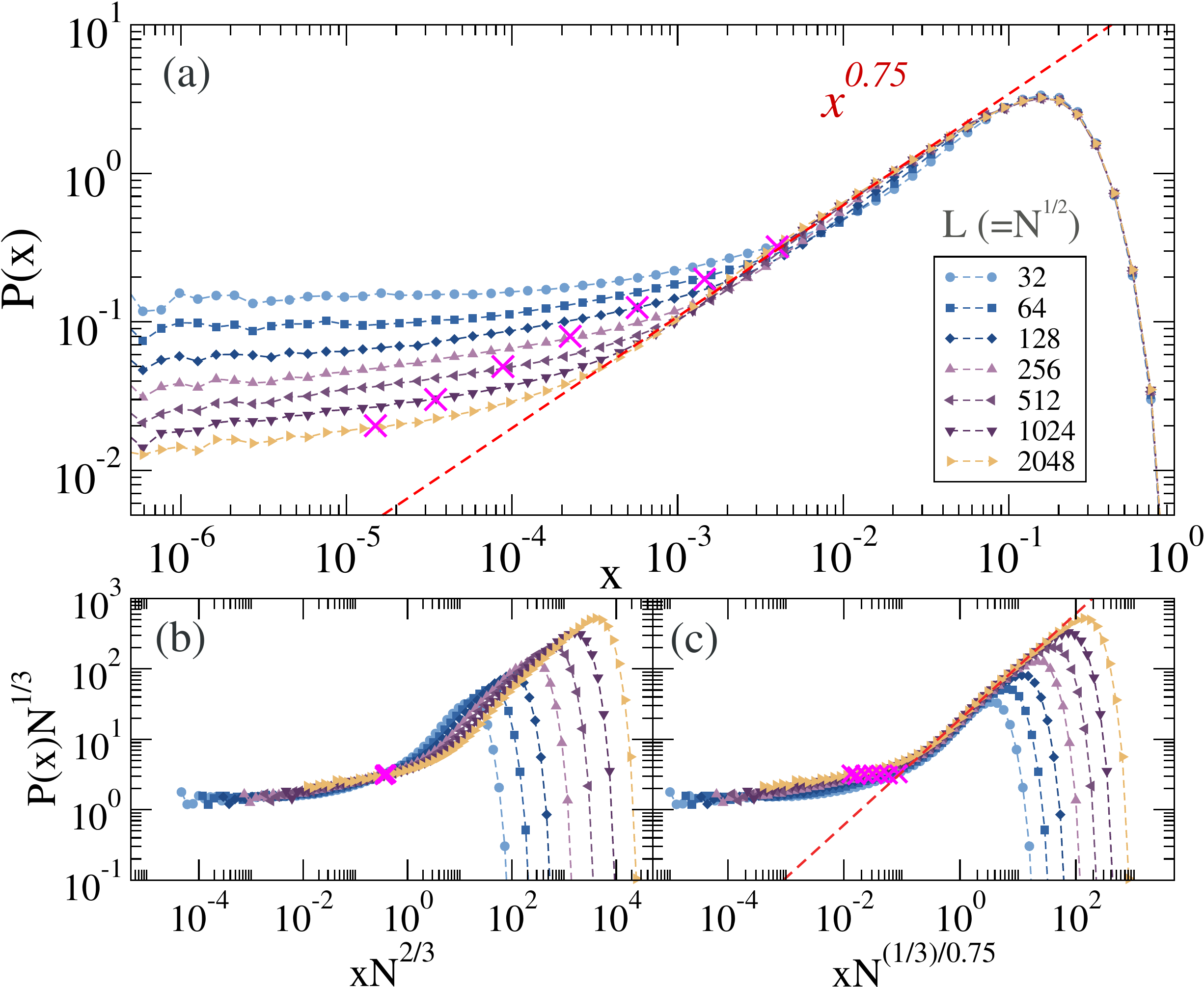} 
\end{center}
\caption{
Distribution of local distances to threshold $P(x)$ in the quasistatic 
driven steady state of Picard's 2D model.
\textbf{(a)} The $P(x)$ distributions. 
Different linear system sizes $L=\sqrt{N}$ are represented with different 
colors/symbols as declared in the label.
Pink crosses indicate the location of $\langle \xmin \rangle$ for the different
system sizes.
\textbf{(b)} $P(x)N^{1/3}$ vs $xN^{2/3}$ testing the scalings of Eqs. 
\ref{eq:xmin_scaling} and \ref{eq:P0_scaling}.
\textbf{(c)} $P(x)N^{1/3}$ vs $x N^{(1/3)/0.75}$ to preserve the
power-law regime $\sim x^\theta$ with $\theta=0.75$ observed in the main 
plot at intermediate values of $x$.
}
\label{fig:PofX_Picard2D}
\end{figure}

\subsection*{Two-dimensional systems (2D)}

\begin{figure}[t!]
\begin{center}
\includegraphics[width=1\columnwidth]{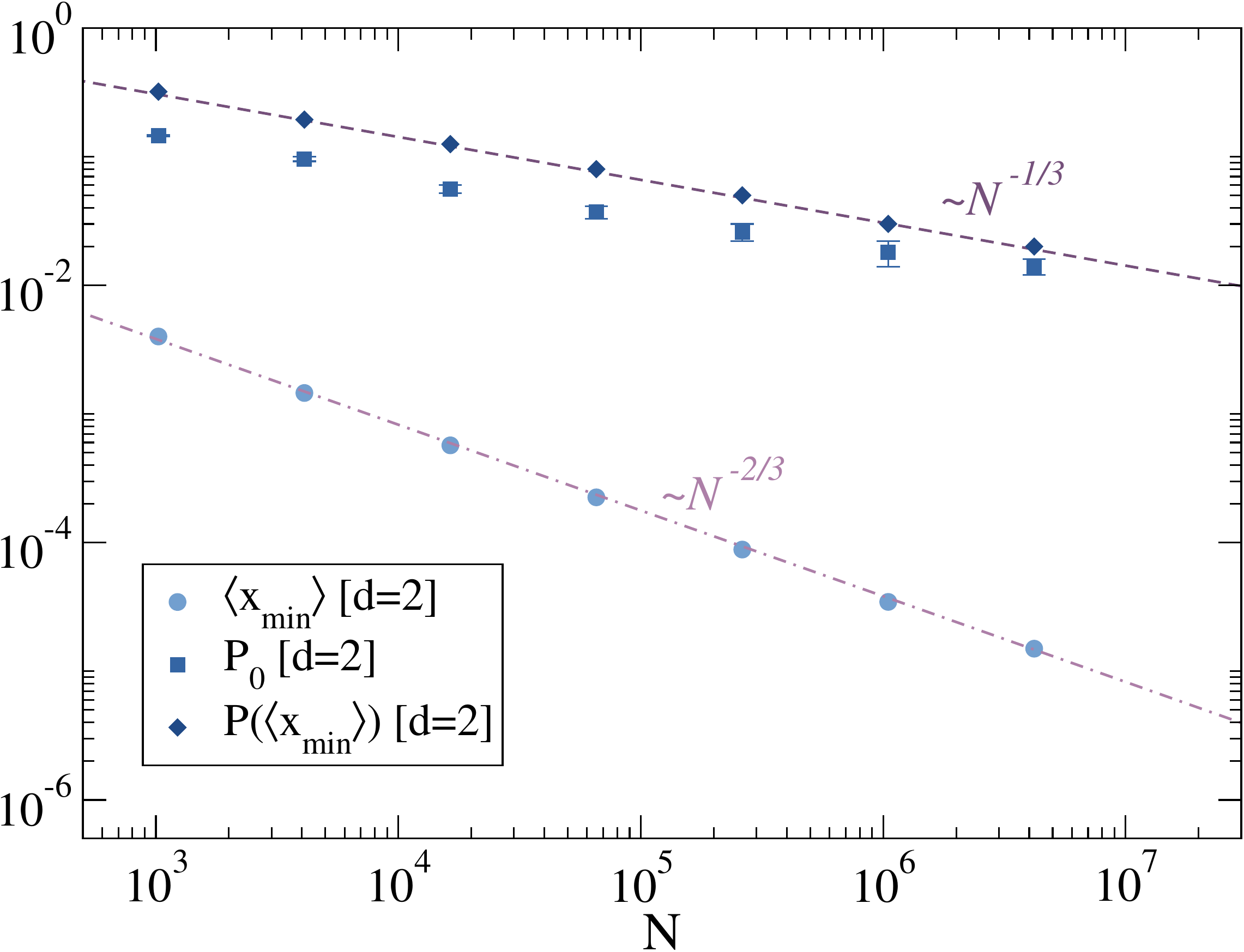} 
\end{center}
\caption{
Dependence of $\langle \xmin \rangle$, $P_0$ and $P(\langle \xmin \rangle)$ with system size $N$ for 2D Picard's model
\label{fig:Picard2D_xmin_and_P0_scaling}.
}
\end{figure}

We start with the $d=2$ case.
Fig. \ref{fig:PofX_Picard2D} shows the distribution $P(x)$ for different system 
sizes $N=L\times L$.
We have collected the values of $x\equiv\Sigma_{{\tt th}}-\Sigma$ 
(see App.~\ref{sec:models} for parameters definitions)
from every block in the system for several configurations in the steady state right 
after an avalanche has finished and before loading the system to the next avalanche.
As discussed in previous sections, $P(x)$ displays --also in this fully
spatial model-- an excess of probability at $x=0$, evidencing the occurrence
of a naturally emerging discrete step for the wandering of the $x$ values. 
Already from the upper panel (Fig.\ref{fig:PofX_Picard2D}(a)) it is evident
the settling of a system-size dependent plateau at $x\to 0$.
This plateau occurs systematically at smaller values of $x$ as $L$ increases.
The form of $P(x)$ has more structure than in the 
random-walk experiments of Sec.~\ref{sec:randomwalks}. 
Now the crossover region between the power-law regime and the plateau 
is broader, the power law range is shrunk due to the natural existence of 
a global drift, and for small systems $P(x)$ even displays an ``S" shape before 
cutting-off when $x$ becomes order 1.
Yet, we can identify for the largest system size a power-law regime
spanning two orders of magnitude in $x$ ($[\sim 8.10^{-4},\sim 8.10^{-2}]$) 
in excellent agreement with $x^\theta$ with $\theta=0.75$ 
(the value expected when $\mu=3/2$ in the discussion of Sec.\ref{sec:mechnoise}).
Let us now check the validity of our predictions in Eqs. \ref{eq:xmin_scaling} 
and \ref{eq:P0_scaling}.
In Fig. \ref{fig:PofX_Picard2D}(b) the same data of panel (a) is plotted as 
$P(x)N^{1/3}$ vs. $xN^{2/3}$. 
The magenta crosses indicate the position $(\xmin,P(\xmin))$ on each
$P(x)$ curve~\footnote{$\xmin$ is independently computed for each system size as 
the arithmetic average of the minimum $x$ values (in the $L\times L$ system) for 
each after-avalanche configuration in the steady state.}.
The coincidence of the horizontal coordinate of these points is the indication that 
Eq.~\ref{eq:xmin_scaling} is very well satisfied. 
According to Eq.~\ref{eq:P0_scaling} we also expect that the plateaus of all curves 
in Fig.~\ref{fig:PofX_Picard2D}(b) level up. 
We see that they do but not perfectly. 
Instead, note that the values of $P(x)$ at $x=\langle \xmin\rangle$ (i.e., 
the vertical coordinate of the crosses) do become
coincident in Fig.~\ref{fig:PofX_Picard2D}(b),
fulfilling better the combination of 
Eqs. \ref{eq:xmin_scaling} and \ref{eq:Pxmin_xmin}
\begin{equation}\label{eq:oneoverN_xmin}
\langle \xmin \rangle P(\langle \xmin \rangle) \sim 1/N .
\end{equation}
Fig. \ref{fig:PofX_Picard2D}(b) is built to display the combined
scaling of $\langle \xmin \rangle$ and $P_0$ (or $P(\langle \xmin \rangle)$). 
If instead we want to get a collapse of the power-law
range of the $P(x)$ distribution for different system sizes, 
we must preserve the power-law exponent in the transformation.
This is done in Fig.~\ref{fig:PofX_Picard2D}(c) where we plot 
$P(x)N^{1/3}$ vs. $xN^{(1/3)/0.75}$, according to the observed
$\theta\simeq 0.75$.
Following our generalized mean-field picture the value $\theta\simeq 0.75$
observed in the 2D elasto-plastic model corresponds
to a mechanical noise with a Hurst exponent $H=\mu^{-1}\simeq 2/3$ 
($\mu=2\theta \simeq 3/2$).
A direct characterization of the mechanical noise to verify this value
was already presented in~\cite{FernandezAguirrePRE2018,FerreroSM2019}, showing
a concurrence of different two-dimensional elasto-plastic models around the
Hurst exponent $H\simeq 2/3$.
Furthermore, very recently compatibility with $\mu\simeq 3/2$ was also
reported in MD simulations~\cite{ruscher2019residual}.

In Fig. \ref{fig:Picard2D_xmin_and_P0_scaling} we show the values of 
$\langle \xmin \rangle$, $P(\langle \xmin \rangle)$ and $P_0$ 
(estimated from the curves in Fig.~\ref{fig:PofX_Picard2D}) 
as a function of $N=L^2$.
Dashed straight lines are displays of the exact power-laws $N^{-2/3}$
and $N^{-1/3}$, not fits.
We can see that the prediction of Eq.~\ref{eq:xmin_scaling} 
work remarkably well and Eq.~\ref{eq:oneoverN_xmin} 
accompanies it perfectly.
The original prediction for the scaling of $P_0$ (Eq.~\ref{eq:P0_scaling})
is also good (as could be seen in the collapses of Figs.~\ref{fig:PofX_Picard2D}(b)-(c)),
but we can also notice that $P_0$ is slowly merging with $P(\langle \xmin \rangle)$
as system size increases, and it is indeed when $N\to \infty$
when we expect them to be equal and Eq.~\ref{eq:P0_scaling} to hold.

\begin{figure}[t!]
\begin{center}
\includegraphics[width=1\columnwidth]{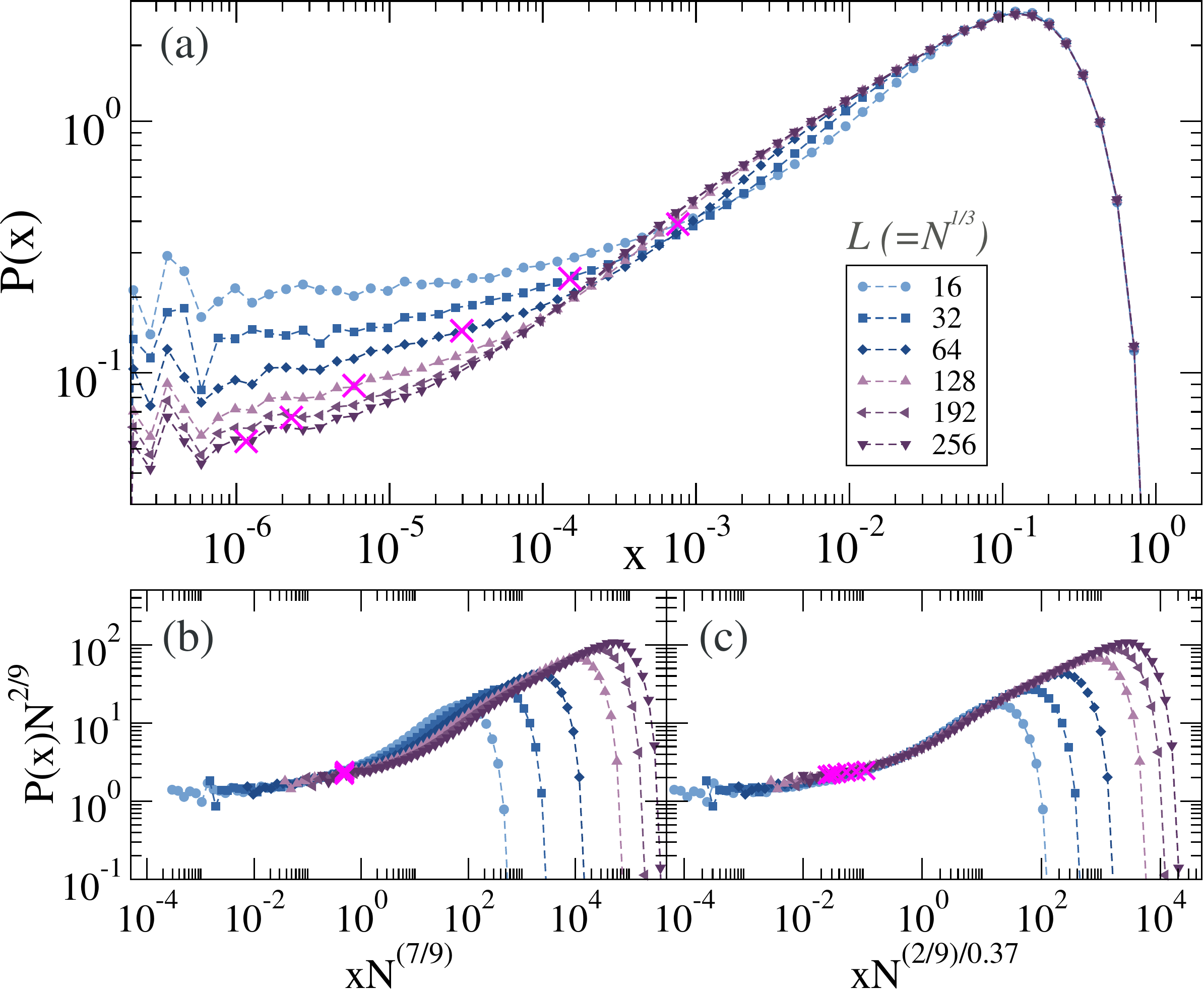} 
\end{center}
\caption{
Distribution of local distances to threshold $P(x)$ in the quasistatic driven
steady state of Picard's 3D model.
\textbf{(a)} The $P(x)$ distributions. 
Different linear system sizes $L=\sqrt[3]{N}$ are represented with different colors/symbols as declared in the label.
Pink crosses indicate the location of $\langle \xmin \rangle$ for the different
system sizes.
\textbf{(b)} $P(x)N^{2/9}$ vs $xN^{2/3}$ testing the scalings of Eqs. 
\ref{eq:xmin_scaling} and \ref{eq:P0_scaling}.
\textbf{(c)} $P(x)N^{2/9}$ vs $x N^{(2/9)/0.37}$ to preserve the
power-law regime $\sim x^\theta$ with $\theta=0.37$ observed in the main 
plot at intermediate values of $x$.
}
\label{fig:PofX_Picard3D}
\end{figure}

\subsection*{Three-dimensional systems (3D)}

Now, let us discuss the three-dimensional case. 
\blue{
Contrary to the 2D case, where the few interaction
kernels that one can choose (corresponding to the different 
kind of volume-preserving applied deformations)
are symmetric under the exchange of $q_x$ and $q_y$,
in 3D the many different possibilities for 
choosing the elastic kernel all are non-symmetric 
respect to the permutations of $q_x$, $q_y$ and $q_z$.
The precise symmetry of the six independent deviatoric modes 
in 3D can be seen for example in~\cite{JaglaPRE2020}.
The results we present here correspond exclusively to the 
kernel shown in Eq. \ref{eq:propagator3d_fourier},
where the way in which the $z$ dimension enters differs from
that of $x$ and $y$.}
In Fig. \ref{fig:PofX_Picard3D} we show
data similar to that in Fig. \ref{fig:PofX_Picard2D} but for the $d=3$ case.
We can fist observe in the raw data of Fig.~\ref{fig:PofX_Picard2D}(a)
that the determination of the $\theta$ exponent is more ambiguous than in $d=2$.
At intermediate values of $x$, say $\sim (0.005-0.1)$, a power-law region can be visualized
and it has an exponent $\theta \simeq 0.35-0.37$, as reported in previous 
works~\cite{LinPNAS2014,liu2015driving}.
Yet, such a value for $\theta$ would imply 
$\mu=2\theta\simeq 0.70 - 0.74 <1$ and therefore $H>1$.
In that case, according to~\cite{LinPRX2016} the drift
becomes dominant and we can't expect the arguments related
to the survival probability of $x$ close to $x=0$ to hold.
Notice nevertheless that, for the largest system sizes, another power-law regime
at smaller $x \sim (10^{-4}-10^{-3})$ is insinuated. 
We will come back on this when discussing systems with different aspect ratios,
but let us advance that such power-law with a steeper slope
would represent a more consistent value for $\theta$ in $d=3$.

\begin{figure}[t!]
\begin{center}
\includegraphics[width=1\columnwidth]{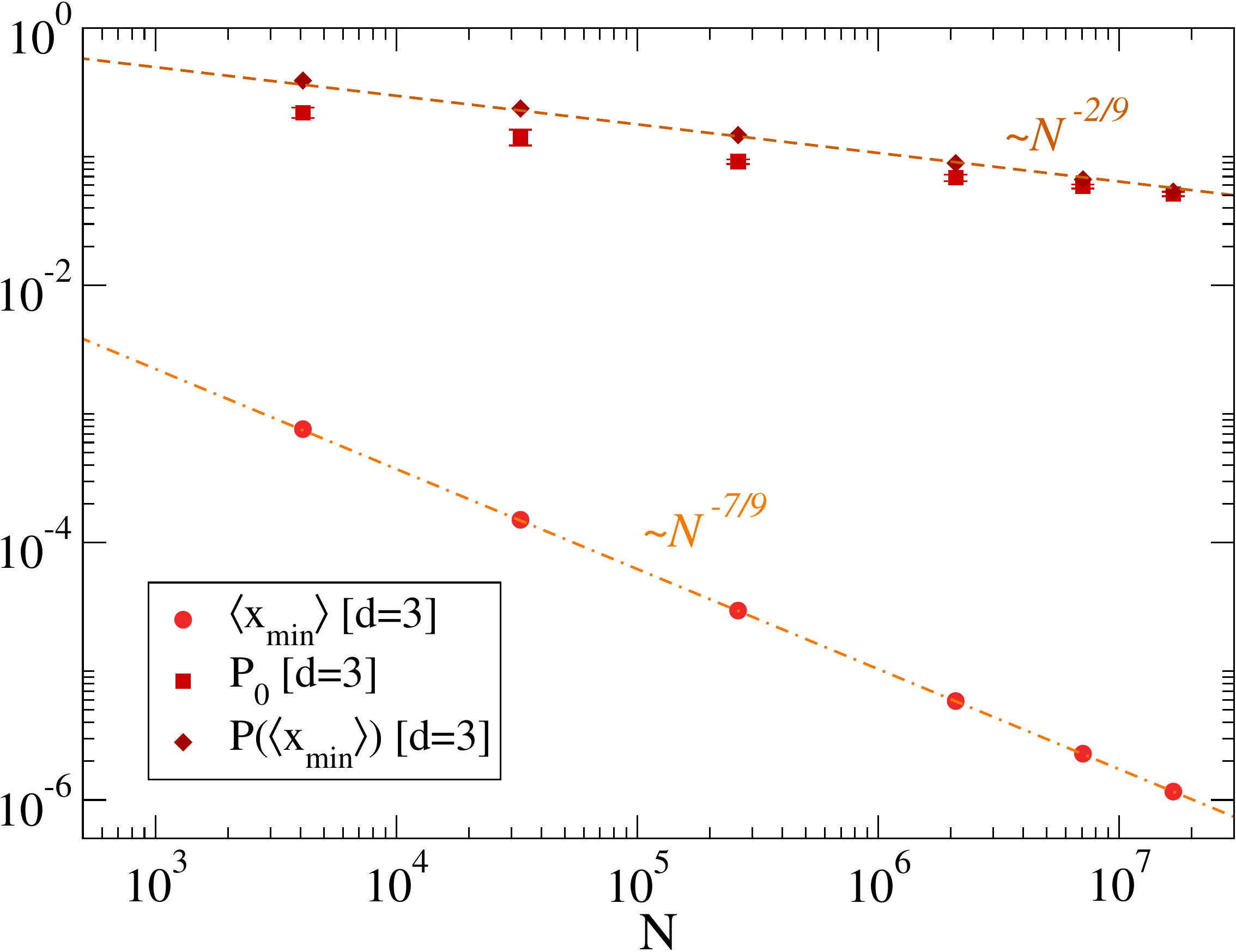} 
\end{center}
\caption{
Dependence of $\langle \xmin \rangle$, $P_0$ and $P(\langle \xmin \rangle)$ with system size $N$ for
(cubic box) 3D Picard's model.
\label{fig:Picard3D_xmin_and_P0_scaling}.
}
\end{figure}

In any case, let us now discuss scalings for the data 
in Fig.~\ref{fig:PofX_Picard3D}.
In Fig. \ref{fig:PofX_Picard3D}(b) we see that the $N$ dependence of 
$\langle \xmin\rangle$ an $P_0$ follows a power-law behavior like
the one predicted by Eqs. \ref{eq:xmin_scaling} and \ref{eq:P0_scaling}
but with clearly different exponents.
Actually, the observed scaling is
$\langle \xmin\rangle \sim N^{-7/9}$ and $P_0 \sim N^{-2/9}$.
Using these values we rescale the $P(x)$ data to obtain 
Fig.~\ref{fig:PofX_Picard3D}(b).
Again, notice that as in the case of $d=2$ the collapse of the 
points $(\langle \xmin\rangle, P(\langle \xmin\rangle))$ (Eqs.~\ref{eq:xmin_scaling} and \ref{eq:Pxmin_xmin}) 
is better than the scaling of the plateaus, which are even hard to define.
If we further consider the power-law regime with an exponent $\theta\simeq 0.37$
we can do as in the $d=2$ case and produce Fig.~\ref{fig:PofX_Picard3D}(c), for 
completeness.

\so{Concerning the predictions of Eqs. \ref{eq:xmin_scaling} and \ref{eq:P0_scaling},
our argumentation in the previous section 
implicitly assumed that all spatial dimensions of the system participate 
on the same footing. 
While the $d=2$ Eshelby propagator (Eq. \ref{eq:propagator2d_fourier}) 
is in fact symmetric against $x$-$y$ exchange, this is not the case for the 
$d=3$ propagator (Eq.\ref{eq:propagator3d_fourier}).}

In Fig. \ref{fig:Picard3D_xmin_and_P0_scaling} we show the values of 
$\langle \xmin \rangle$, $P(\langle \xmin \rangle)$ and 
$P_0$ (estimated from Fig. \ref{fig:PofX_Picard3D})
as a function of $N=L^3$ for $d=3$.
Dashed straight lines simply display the power laws
$\sim N^{-2/9}$ and $\sim N^{-7/9}$, they are not fits. The measured values shown in
Fig. \ref{fig:Picard3D_xmin_and_P0_scaling} follow these trends very well.
\blue{These values do not coincide with the predictions of Eqs. \ref{eq:xmin_scaling} and \ref{eq:P0_scaling}.
We believe the main reason is that our argumentation in the previous section 
implicitly assumed that all spatial dimensions of the system participate 
on the same footing. 
As we already stressed it, while the $d=2$ Eshelby propagator (Eq. \ref{eq:propagator2d_fourier}) 
is in fact symmetric against  exchange of axis, this is not the case for the 
$d=3$ propagator (Eq.\ref{eq:propagator3d_fourier}).}

\begin{figure}[t!]
\begin{center}
\includegraphics[width=1\columnwidth]{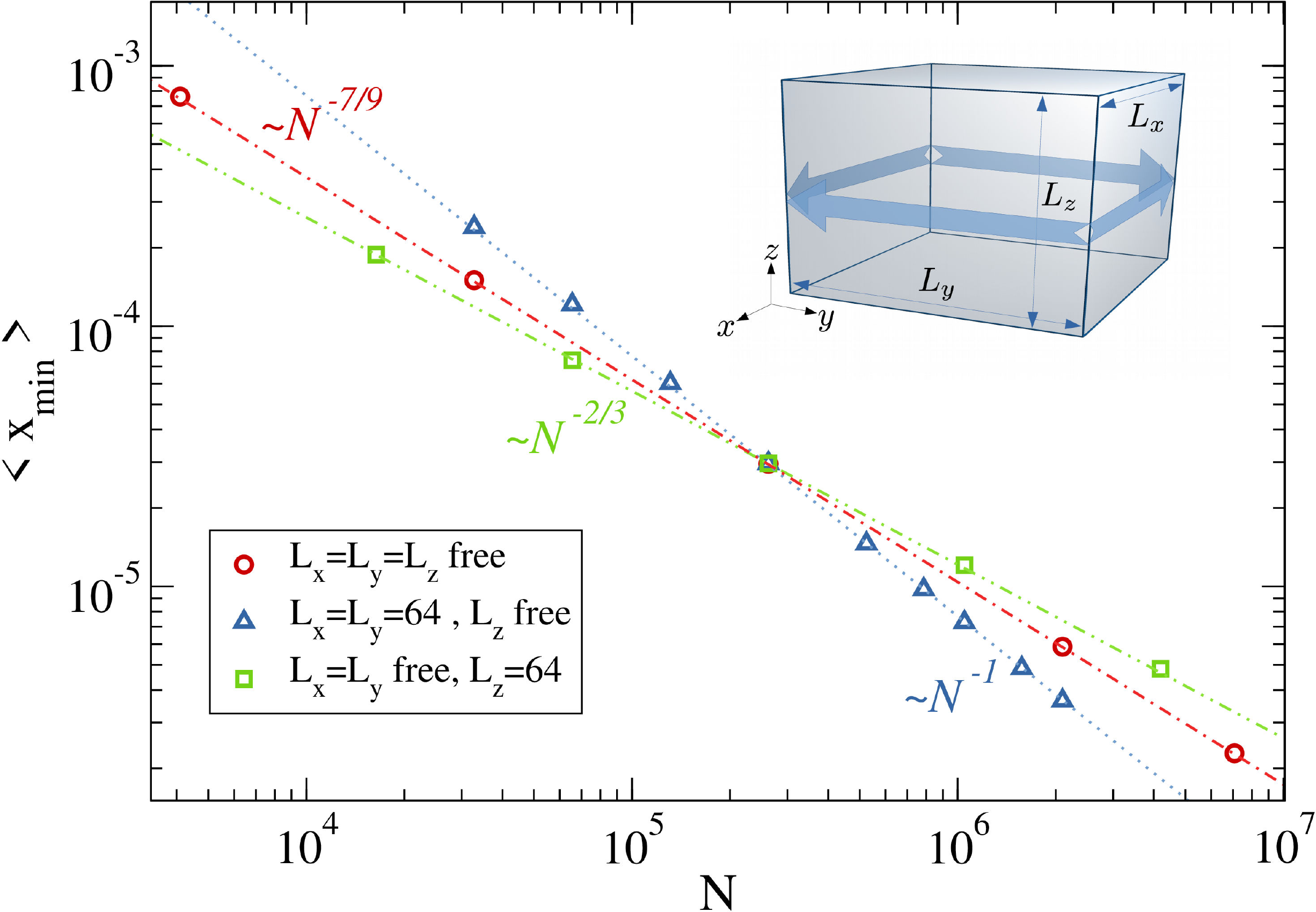} 
\end{center}
\caption{\label{fig:av_xmin_3Dassymetric}
Dependence of $\xmin$ with system size $N$ for the 3D Picard's model
with different aspect ratios.
The sketch of the inset represents the 3D simulation box with its
dimensions $L_x$, $L_y$ and $L_z$, and arrows indicating the
strain deformation that gives rise to the propagator that we 
use for $d=3$ in this work (Eq.\ref{eq:propagator3d_fourier}).
}
\end{figure}

\blue{We can provide a partial explanation for the values found for the N dependence of 
$\langle\xmin\rangle$ and $P(\langle \xmin \rangle)$ (or $P_0$) in 3D in the following way}
\so{These, so-far empirical, values for the scaling of
$\langle\xmin\rangle$ and $P(\langle \xmin \rangle)$ (or $P_0$)
can be rationalized in the following way.}
\blue{First of all, notice that for $q_z=0$ the three dimensional kernel (Eq. \ref{eq:propagator3d_fourier}) reduces to the
two dimensional one (Eq. \ref{eq:propagator2d_fourier}).}
\so{Let us consider}\blue{We will make the assumption} that the non-trivial scaling of $\langle \xmin \rangle$ 
is still governed by the finite-kick walk analysis that we did in 
Sec. \ref{sec:mechnoise}, but \so{we add to that} \blue{in which the $z$ coordinate has to be treated as}  a `dumb' independent 
dimension.
This is, let's think on the $d=3$ case as a collection of several $d=2$
systems \blue{stacked in the $z$ direction, and} evolving in parallel. 
If we take, $L_z$ systems of size $L\times L$ and choose after each avalanche
the minimum $x$ among all of them, we would have a $\langle \xmin \rangle$ scaling as
\begin{equation}
\langle \xmin \rangle \sim L^{-4/3}L_z^{-1}
\label{eq:xmin3d}
\end{equation}
and 
\begin{equation}
P_0 \sim L^{-2/3}
\label{eq:p03d}
\end{equation}
(note that $P_0$ turns out to be independent of $L_z$).
When $L_z=L$ this leads to the scaling $\langle \xmin \rangle \sim N^{-7/9}$ and $P_0\sim N^{-2/9}$
(with $N=L^3$) that we observe in Fig. \ref{fig:Picard2D_xmin_and_P0_scaling}.
In fact, simulations in systems with different $L_x=L_y=L$ and $L_z$ 
show that Eqs.~\ref{eq:xmin3d} and \ref{eq:p03d} are very well satisfied,
as we will see in the following.

\begin{figure}[t!]
\begin{center}
\includegraphics[width=1\columnwidth]{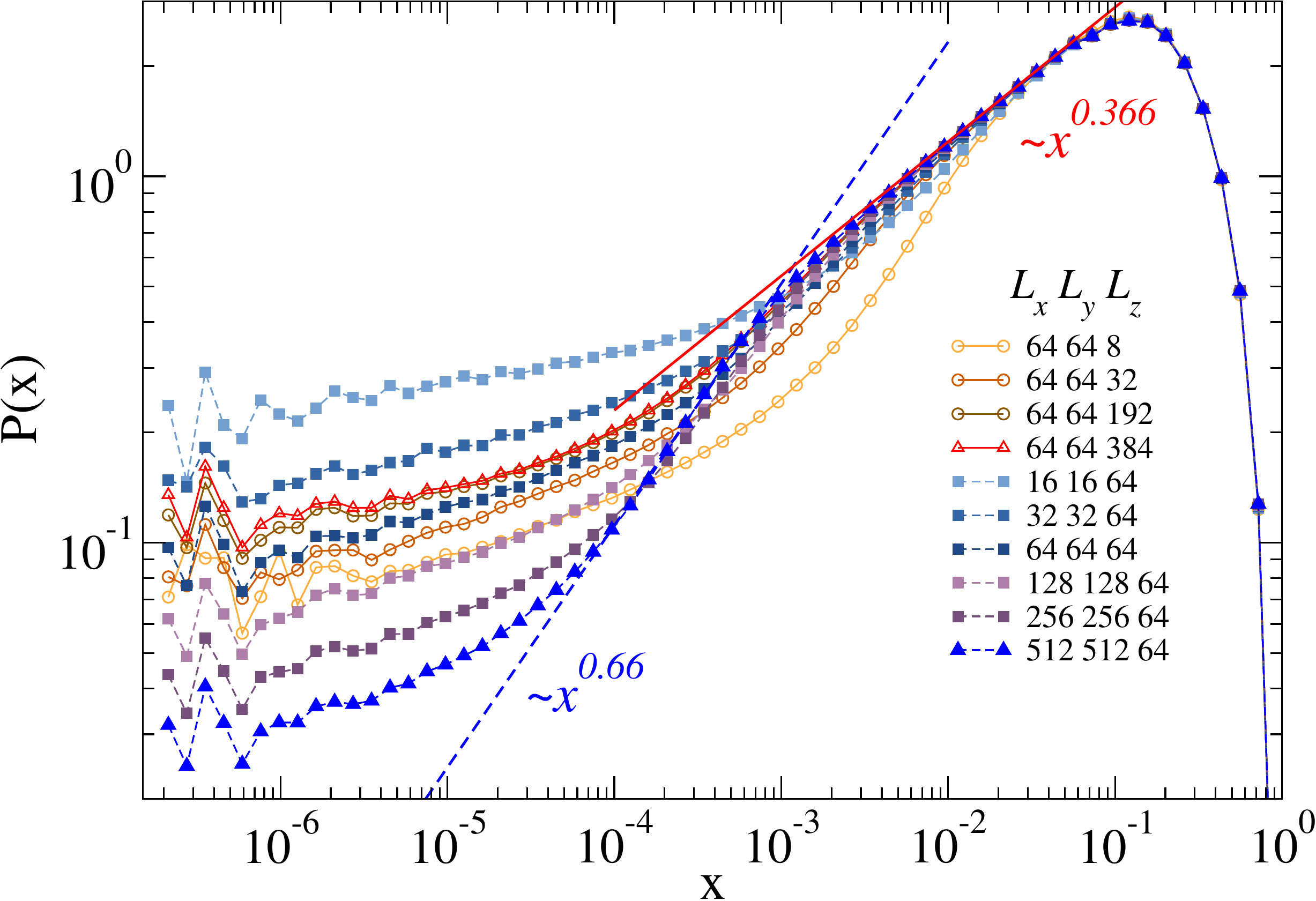} 
\end{center}
\caption{
$P(x)$ for Picard's 3D model with different aspect ratios.
For the largest system size of each size grow scheme 
(varying $L_x=L_y$ or varying $L_z$), the power-law regime
visible at the smallest values of $x$ is marked by a straight 
line to guide the eye.
}
\label{fig:PofX_Picard3D_aspectratios}
\end{figure}

Fig.~\ref{fig:av_xmin_3Dassymetric} shows the scaling of $\langle\xmin\rangle$ 
for different cases.
First, the $N=L^3$ case is reproduced from Fig.~\ref{fig:Picard3D_xmin_and_P0_scaling}
for comparison.
Then, we increase the system size while fixing $L_x=L_y$ and varying only $L_z$ 
(the dimension perpendicular to the shear plane, 
that enters in a `different' way than the other two in the
propagator~\ref{eq:propagator3d_fourier}).
This yields a scaling $\langle\xmin\rangle \sim N^{-1}$  controlled by 
$\langle\xmin\rangle \sim L_z^{-1}$ since the system size in the other 
two dimensions is fixed.
Finally, we do inversely and we increase the system size by growing $L_x=L_y$ 
and keeping $L_z$ fixed.
This yields a scaling $\langle\xmin\rangle \sim N^{-2/3}$ controlled by 
a scaling of $\langle\xmin\rangle \sim L^{-4/3}$ (eq.~\ref{eq:xmin3d})
for both $L_x$ and $L_y$.
Notice that when the system size is increased in this way 
(at a fix perpendicular direction to the shear plane) we recover
the scaling observed in the MD simulations of~\cite{LernerPRE2009,Karmakar2010,KarmakarPRE2010b},
that shows no exponent difference between $d=2$ and $d=3$.

In Fig.~\ref{fig:PofX_Picard3D_aspectratios} we take a look to 
the $P(x)$ distributions in these asymmetric boxes for different 
aspect ratios.
On one hand, we have fixed $L_x=L_y=64$ and vary $L_z$ between 8 and 384.
On the other hand, we have fixed $L_z=64$ and vary $L_x=L_y$ between 16 and 512.
Notice first that, when $L_z$ is the only changing dimension, the plateau
level actually increases, with a small positive power, 
and it seems to saturate for large sizes around $P_0\sim 0.12$.
So, the strong $\langle\xmin\rangle$ scaling decreasing as $1/N$, 
is accompanied by a barely changing $P_0$ with $N$, as we could have expected
from Eq.~\ref{eq:P0_xmin}.
These curves for $P(x)$ have the particularity that they only show a  
power-law regime at `large' values of $x$, and they correspond to an 
`abnormally small' value of $\theta$,
coincident with the many times reported~\cite{LinPNAS2014,liu2015driving} 
but never truly justified $\theta\simeq 0.35-0.37$ in 3D.
This $\theta$ value would point to $\mu<1$, 
beyond the assumptions used for the derivation of our scaling arguments.

Now, if we analyze the curves when varying $L_x=L_y$ at fix $L_z$ things
change dramatically.
First, the plateau level is `well behaved'' decreasing as $N$ increases.
In fact, a reasonable $P_0 \sim N^{-0.3}$ accompanies the scaling of $\xmin$
shown in Fig.\ref{fig:av_xmin_3Dassymetric} for this case.
Secondly, the larger system sizes clearly display a different power-law
at intermediate $x$ values, with $P(x)\sim x^{0.66}$ in such range.
As it stands closer to the boundary $x=0$, that will be the power-law
dominating the system's dynamics close to the transition
(e.g., the value of the flowcurve exponent $\beta$~\cite{FerreroSM2019,FerreroPRL2019}).
In fact, in $d=3$, $H=1/(2\theta) \simeq 0.75-0.77$ is expected~\cite{FerreroPRL2019},
consistent with $\theta \simeq 0.65-0.66$. 
Moreover, $\theta\simeq 0.66$ suggests a value of $\mu=2\theta\simeq 1.33$
in $d=3$, which brings the problem back into the range of validity of our
general assumptions for the derivation of the scalings (\ref{eq:xmin_scaling} 
and \ref{eq:P0_scaling}).

\so{How this} 
\blue{It needs to be stressed that the occurrence of these} 
two clearly different finite size scalings in 3D -- 
\textit{(i)} growing the system in the direction perpendicular to the shear plane or
\textit{(ii)} growing the system in the directions of the shear plane --
\so{are combined to produce the empirically observed $\langle\xmin\rangle\sim N^{-7/9}$
and $P_0\sim N^{-2/9}$ for 3D cubic boxes is an open question.}
\blue{remains as an open issue (see discussion below).}

\section{Summary and discussion}
\label{sec:conclusions}
\vspace{-0.2cm}

In this paper we have considered the problem of the strain load 
$\Delta\gamma$ needed to trigger consecutive avalanches 
in the steady state of quasistatically deformed amorphous solids.
In particular, we studied the finite-size scaling of its mean 
value $\langle\Delta\gamma\rangle$.
The values of $\Delta\gamma$ are intimately related to the 
distribution $P(x)$ of local distances to instability $x$; 
$\langle\Delta\gamma\rangle$ is simply proportional to 
the average value of the minimum $x$ across the system, 
namely $\langle\Delta\gamma\rangle\sim \langle\xmin\rangle$.
We have built a theoretical argument starting by simple random walks
of $x$ with an absorbing boundary to show how the effect of a 
\textit{discrete} step induces a finite value of $P(x)$ at the boundary.
Then we stood on an alternative mean-field modeling approach for the 
yielding phenomena~\cite{FernandezAguirrePRE2018,FerreroSM2019}, 
considering as the physically relevant case the one in which the mechanical 
noise is generated by extended and collective plastic events, leading to a 
fat tail noise distribution with $1\!<\!\mu\!<\!2$.
The mechanical noise generated by these avalanches has \blue{indeed} a 
discrete nature, and therefore the distribution of $P(x)$ is expected to 
acquire a finite value as $x\to 0$, namely $P(x\to 0)=P_0\ne 0$.
\blue{More importantly, the discreetness in the mechanical kicks is not 
the trivial $\sim 1/N$ finite-system effect, but one that has to do also
with the mean avalanche size in the system (e.g., see Eq.~\ref{eq:mdsigma}).
This holds for any $1\!<\!\mu\!<\!2$ and we explicitly derive the $P_0$ vs 
$N$ scaling in that case.
}
\so{This}\blue{The} scenario is confirmed by extensive numerical simulations of a classical 
elastoplastic model in 2 and 3 dimensions.

Even though the value of $P_0$ decreases to zero as $N\!\to\!\infty$,
and therefore it could be naively considered a finite-size effect, 
its behavior with system size happens to be precisely what governs 
the scaling of $\langle\xmin\rangle$, and thus of $\langle\Delta\gamma\rangle$, 
our quantity of interest.
Our theoretical analysis 
is able to justify a universal dependence $\langle\Delta\gamma\rangle \sim N^{-\alpha}$, 
with $\alpha=2/3$, independent of spatial dimension and system parameters, 
as is actually found in MD simulations~\cite{LernerPRE2009,Karmakar2010}.
Moreover, we have no need to assume a particular shape for the 
energy barriers~\cite{Karmakar2010} in doing so.
It is worth mentioning nevertheless, that, as most of the numerical 
literature on the field, our construction assumes so far an athermal system.  
In this case the dynamics is dominated by the minimal value of distance to 
instability, $\xmin$, at every loading step.
A finite temperature in a thermodynamic system ($N\to\infty$) may
blur this (otherwise strictly) extremal dynamics.
It might be an interesting problem for future works to analyze 
how our predictions are impacted by a finite temperature.

In the numerical results presented here for EP models in $d=2$ the 
value $\alpha=2/3$ is clearly obtained.
However, the corresponding results in symmetric (i.e., cubic) $d=3$ 
systems display a different value $\alpha\simeq 7/9$. 
We have identified a possible reason for this discrepancy in $d=3$
in an unforeseen $\langle\xmin\rangle$ scaling dependence with the 
linear size of the sample along different directions relative to the 
externally applied shear.
In contrast with the $d=2$ case, the interaction kernel in $d=3$
is not symmetric in all coordinates. 
Growing the system in the direction perpendicular to the shear plane
has an effect markedly different on $\langle\xmin\rangle$,
than growing it in the other directions.
By studying $d=3$ systems of different aspect ratio we addressed these
multiple scalings, showing that in the case in which 
the dimension perpendicular to the shear plane is kept fixed,
$\alpha\simeq 2/3$ is recovered even in $d=3$ elasto-plastic systems.
The different scalings of $\langle\xmin\rangle$ and $P_0$ with the different 
linear dimensions of a 3D system can be rationalized by a gedanken problem
in which the 3D system behaves as a collection of independent 2D systems.
However, there is no basis to expect that this is actually the way in which 
a 3D system behaves and we know that the dynamics of interactions is more 
complex than that. 
\blue{
The kernel asymmetry might be a weakness of the EP simplification 
and non-physically dominant at the end of the day.
This is what one could interpret from the fact that classical MD results~\cite{LernerPRE2009,Karmakar2010,KarmakarPRE2010b}
maintain the $\langle\xmin\rangle\sim N^{-2/3}$ scaling.
Moreover, a recent proposal of `augmented' elastoplastic models
tries to incorporate (among other things) the fact that
shear strain in {\it any} direction due to a rearrangement can 
trigger the next rearrangement equally well.
Successive rearrangements observed in MD are ``isotropically 
distributed'' and not concentrated in the strain direction
prescribed by the imposed deformation~\cite{zhang2020interplay}.
If the interaction kernel is symmetrized somehow 
our predictions Eqs.~\ref{eq:xmin_scaling} and \ref{eq:Pxmin_xmin},
turn to be valid in the elastoplastic 3D case as well.
We have checked this so far for synthetic, non-physical,
kernels only (data not shown).
}

In any case, a definite value of $\alpha$ implies additional predictions 
on other critical exponents of the yielding transition. 
For instance, the avalanche distribution exponent 
$\tau$ and the fractal dimension of avalanches 
$d_f$ are linked to $\alpha$ through ~\cite{LinPNAS2014,FerreroSM2019}
\begin{equation}\label{eq:dalpha}
d\alpha=d-d_f(2-\tau). 
\end{equation}
(note that this relation is usually written using $\theta$ 
instead of $\alpha$, by applying the extra assumption 
$\alpha=1/(1+\theta)$, that we consider not justified 
in the steady state).
A unique value $\alpha=2/3$ in $d=2$ implies $d_f(2-\tau)=2/3$. 
Most of the values reported in the literature satisfy this relation. 
In particular, we have tested for six different EP models~\cite{FerreroSM2019}
$d_f\simeq 1$ and $\tau\simeq 1.33$.  
For $d=3$, we must still understand which is the value of $\alpha$
that we should expect, but $d_f$ and $\tau$ could also suffer from
an asymmetry effect if Eq.~\ref{eq:dalpha} is expected to hold.

Finally, all this picture should be compatible with known results 
for the as-quenched state; with rigorous power-laws for $P(\xmin)$ and 
$P(x)$ at small arguments.
We believe that the effective mechanical noise governing the distribution
$P(x)$ and its properties, like the one that defines the finite-size 
scaling of $\langle\xmin\rangle$, must display systematic biases
in the non-universal transient.
While we will not venture to link transient values of $\mu$ (or $H$) with $\theta$
in such a regime (which, furthermore, is only measurable on a given system
size for certain ranges of initial annealing), our guess is that avalanches 
progressively build up and their geometry 
-encoded in $d_f$~\cite{LinPNAS2014,lin2015criticality}- varies with 
strain, therefore modifying the effective noise, until it reaches a 
steady distribution governed by $1<\mu<2$.

\subsection*{Conclusion}

In conclusion, we have provided a novel interpretation of the finite size 
scaling of $\langle\Delta\gamma\rangle$ in the steady state of 
amorphous systems under deformation.
This interpretation seems to conciliate MD simulation results and 
EP constructions, otherwise in contradiction in this limit.
While the hypothesis of a marginal stability behavior, rooted
in the celebrated $P(x)\sim x^\theta$ pseudo-gap, has been proved
to hold in the as-quenched isotropic state of model 
glasses~\cite{LernerPRE2009, Karmakar2010} and still renders 
important outcomes in the transient~\cite{lin2015criticality,ShangPNAS2020}, 
it does not seem to apply `as-is' to the steady state case.
There, at least, the system dynamics is correlated at the level of 
avalanches and this naturally produces a finite value of $P(x)$ as 
$x\to 0$, when observing the $P(x)$ distribution in the quasistatic 
limit, justified on a discrete step for the effective dynamics of 
the $x$ values.
This behavior of $P(x)$ does not invalidate the essence
of the yielding transition, anchored in the sub-extensive scaling of 
$\langle\Delta\gamma\rangle^{-1}$; since the level of such asymptotic 
plateau at small $x$ is itself dependent on $N$ and is shown to govern 
the behavior of
$\langle\Delta\gamma\rangle\sim N^{-\alpha}$, independently on $\theta$.

Some questions remain open, and we hope they will motivate further
endeavors on the subject.
But we believe that this is a first step in shedding light on a
probable misconstruction in the field, based in a wrong extrapolation 
of arguments valid in the early deformation regime to the steady state case.

\section*{Acknowledgments}

We are indebted to D. Vandembroucq and C. Maloney 
for illuminating discussions on an early version of this manuscript.
We sincerely thank the critical feedback provided by M. Wyart
on a first draft of this work.
We also acknowledge exchanges with J.-L. Barrat, E. Lerner, 
J. Rottler and B. Tyukodi.
EEF acknowledges support from PICT-2017-1202.

\appendix

\section{Elastoplastic model and simulation protocol}
\label{sec:models}

EP models are intended to describe amorphous materials at a coarse-grained-level,
laying in between the particle-based simulations and the
continuum-level description~\cite{NicolasRMP2018}.
In short, the amorphous solid is represented by a coarse-grained scalar
stress field $\Sigma(\br,t)$, at spatial position $\br$ and time $t$ under
an externally applied shear strain. 
Space is discretized in blocks (e.g., square lattice).
At a given time, each block can be ``inactive'' or ``active''
(i.e., yielding).
This state is defined by the value of an additional variable:
$n(\br,t)=0$ (inactive), or $n(\br,t)=1$ (active).
An over-damped dynamics is imposed for the stress on each block, following some
basic rules:
(i) The stress loads locally in an elastic manner while the block is inactive.
(ii) When the local stress overcomes a local yield stress, a \textit{plastic event} 
occurs with a given probability, and the block becomes ``active'' ($n(\br)$ is set to one).
Upon activation, dissipation occurs locally, and this is expressed as a progressive 
drop of the local stress, together with a redistribution of the stresses in the rest
of the system in the form of a long-range elastic perturbation.
A block ceases to be active when a prescribed criterion is met.
The auxiliary binary field $n(\br, t)$ shows up in the equation of motion for the
local stress $\Sigma(\br,t)$, defining a dynamics that is typically non-Markovian.
While the structure of the equation of motion for the local stresses is almost unique
in the literature, both its parameters and the rules governing the transitions of
$n(\br)$ ($0\rightleftharpoons 1$) show a variety of choices. 

We define our EP model as a $d$-dimensional scalar field $\Sigma(\br,t)$, 
with tipically $d=2$ or 3, and $\br$ discretized on a square/cubic lattice and each block $\Sigma_i$ subject
to the following evolution in real space
\begin{equation}\label{eq:eqofmotion1}
\frac{\partial \Sigma_i(t)}{\partial t} =
  \mu\dot{\gamma}^{\tt ext}  +\sum_{j} G_{ij} n_j(t)\frac{\Sigma_j(t)}{\tau} ;
\end{equation}
where $\dot{\gamma}^{\tt ext}$ is the externally applied strain rate,  and the kernel $G_{ij}$ is the Eshelby
stress propagator~\cite{Picard2004}.

It is sometimes convenient to explicitly separate the $i=j$ term in the previous sum, as
\begin{equation}\label{eq:propagator2d_fourier}
\frac{\partial \Sigma_i(t)}{\partial t} =
  \mu\dot{\gamma}^{\tt ext}  - g_0 n_i(t)\frac{\Sigma_i(t)}{\tau} + \sum_{j\neq i} G_{ij} n_j(t)\frac{\Sigma_j(t)}{\tau} ;
\end{equation}
where $g_0\equiv -G_{ii} > 0$ (no sum) sets the local stress dissipation rate for
an active site.
The form of $G$ is $G(\br,\br') \equiv G(r,\varphi)\sim\frac{1}{\pi r^2}\cos(4\varphi)$ in polar coordinates,
where $\varphi \equiv \arccos((\br-\br')\cdot\br_{\dot{\gamma}^{\tt (ext)}})$ and
$r \equiv \left|\br-\br'\right|$. For our simulations we obtain $G_{ij}$ from the values of the propagator in Fourier space $G_{\bf q}$, defined as
\begin{equation}
G_{\bf q} = -\frac{4q_x^2q_y^2}{(q_x^2+q_y^2)^2}.
\label{eshelby_kernel}
\end{equation}
for $\bf q\ne 0$ and 
\begin{equation}
G_{\bf q=0}=-\kappa
\label{eshelby_kernel_q0}
\end{equation}
with $\kappa$ a numerical constant (see below).
Note that in our square numerical mesh of
size $L\times L$, $q_x^2$, $q_y^2$ must be understood as 
\begin{equation}
q_{x,y}^2\equiv 2-2\cos\left (\frac{\pi m_{x,y}}{L}\right )
\end{equation}
with $m_{x,y}=0,..., L-1$.

The elastic (e.g. shear) modulus $\mu=1$ defines the stress unit, and the mechanical
relaxation time $\tau=1$, the time unit of the problem.
The last term of (\ref{eq:propagator2d_fourier}) constitutes a \textit{mechanical noise}
acting on $\Sigma_i$ due to the instantaneous integrated plastic activity
over all other blocks ($j\neq i$) in the system.

The picture is completed by a dynamical law for the local state variable 
$n_i=\{0,1\}$. 
We define hereafter the rule corresponding to the Picard's model \cite{Picard2005} 
that we use:

\begin{equation}\label{eq:rulesPicard}
n_i : \begin{cases} 0 \rightarrow 1 & \mbox{at rate~} \tau_{\tt on}^{-1} \mbox{\quad if \quad} \Sigma_i >\Sigma_{{\tt th}i} \\
                    0 \leftarrow 1  & \mbox{at rate~} \tau_{\tt off}^{-1} 
      \end{cases}
\end{equation}
\noindent where $\tau_{\tt on}$ and $\tau_{\tt off}$ are parameters and $P(\Sigma_{{\tt th}i})=\delta(\Sigma_{{\tt th}i}-1)$.

In $d=3$, the Eshelby kernel for one scalar component of the deviatoric strain 
in Fourier space can be written as
\begin{equation}\label{eq:propagator3d_fourier}
G_{\bf q}^{\tt 3D} = -\frac{4q_x^2q_y^2+q_z^2(q_x^2+q_y^2+q_z^2)}{(q_x^2+q_y^2+q_z^2)^2}
\end{equation}
and the dimensional extension of the dynamics is straightforward.

\subsection{Quasistatic protocol}

For the analysis of avalanche statistics, it is convenient to have a
protocol that allows for the triggering and unperturbed evolution
(no driving) of avalanches until they stop, 
guaranteed by a degree of stress non-conservation $\kappa>0$
(we use $\kappa=1$, as in previous strain-controlled EP models implementations~\cite{Martens2012,nicolas2014rheology,liu2015driving},
unless otherwise specified).
This is the quasi-static protocol described here.

Starting from any stable configuration, i.e., no
site is active and no site stress is above its local
threshold ($n_i=0$ and $\Sigma_i<\Sigma_{{\tt th}i}$ for all sites),
the next avalanche of plastic activity is triggered by
globally increasing the stress by the minimum amount necessary
for a site to reach its local threshold.
That site (the weakest) is activated at threshold with no stochastic delays;
it perturbs the stress values of other sites and the rest of the avalanche evolves
without any external drive following the dynamics prescribed
by Eq. (\ref{eq:propagator2d_fourier}) (and the corresponding activation rule)
with $\gdot=0$. 
The avalanche stops once there are no more active sites and all stresses
are below their corresponding thresholds again.
At this point the loading process is repeated.
For each simulation run, data is collected only in the steady-state.

\section{Model with a quenched random kernel}
\label{sec:random_k}

\begin{figure}[t!]
\begin{center}
\includegraphics[width=1\columnwidth]{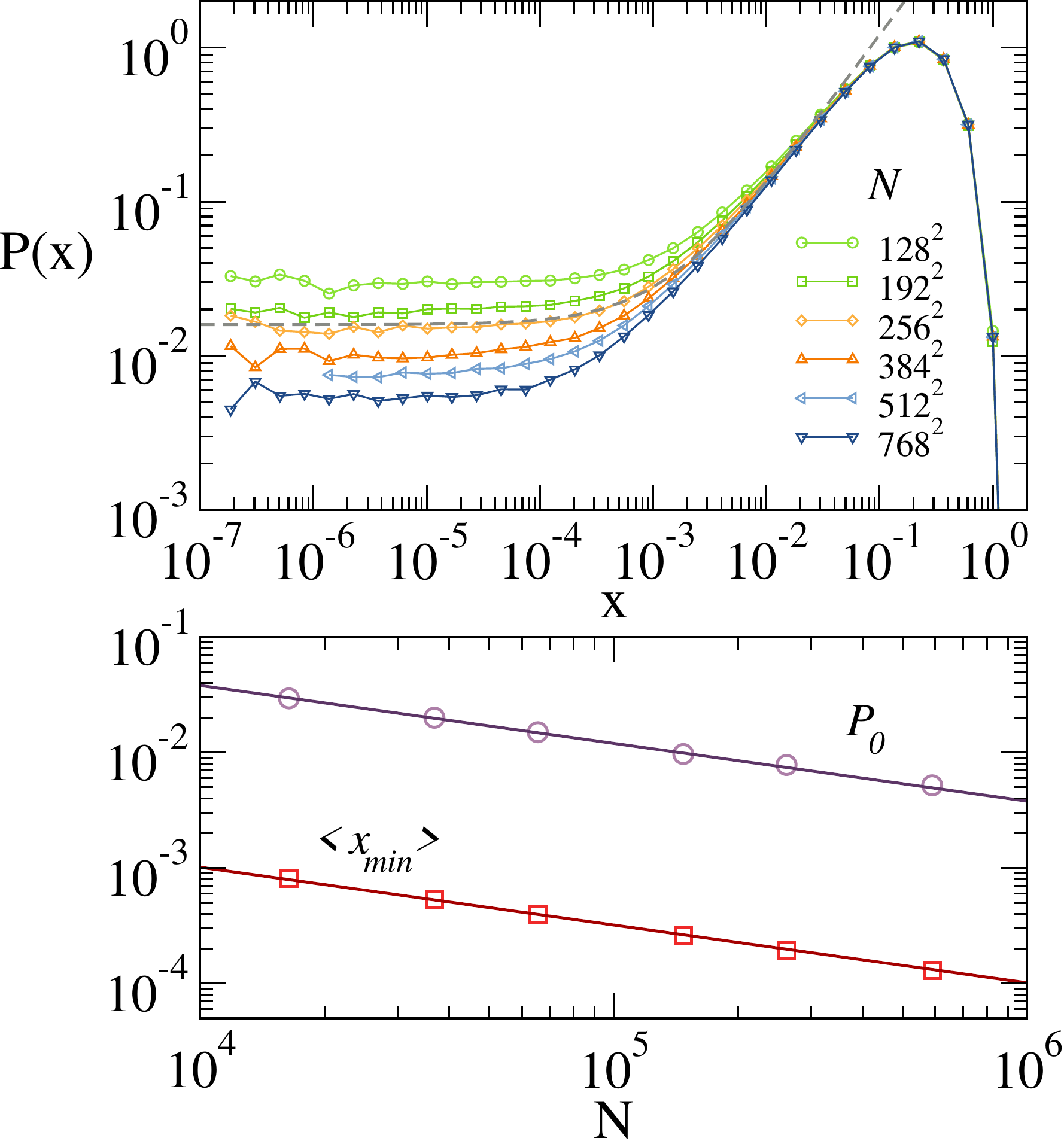} 
\end{center}
\caption{
(a) The form of $P(x)$ for system with different number of sites $N$ in 
the quenched random kernel case. 
The dashed line displays the expected behavior $P(x)= C_1+C_2x$ on the 
$N=256^2$ data. 
(b) The scaling of $P_0$ and $\langle \xmin\rangle$ with $N$. 
Symbols are the result of simulations.
Straight lines indicate the expected $\sim N^{-1/2}$ dependence.
}
\label{fig:random}
\end{figure}

In  this  Section  we  analyze  the  properties of a model
with a different form of the interaction kernel.  
Instead of using the appropriate interaction to describe the properties of yielding,
namely the Eshelby kernel presented in Eq.~\ref{eshelby_kernel}, we consider a model in 
which the $G_{\bf q}$ kernel takes random values.
In concrete, we use 
\begin{equation}
G_{\bf q} = -\mbox {RND}({\bf q}).
\label{random_kernel}
\end{equation}
where $\mbox {RND}({\bf q})$ stands for an independent random number chosen from 
a flat distribution between 0 and 1 for each value of ${\bf q}$.
Note that this is a ``quenched" random kernel, since the form of $G_{\bf q}$
is chosen once and for all at the beginning of the simulation. 

Although this is probably not a realistic model to describe any physical situation, 
there are a few reasons that make the study of this model interesting. 
The first one concerns its relation with another version of a ``random" 
yielding model, namely the Hébraud-Lequeux (HL) model~\cite{Hebraud1998,AgoritsasEPJE2015}. 
In its essence, the HL model for a system with $N$ sites considers that every 
time a single site performs a plastic re-accommodation, it produces a random 
kick of finite variance $\sigma$ (with $\sigma\sim N^{-1/2}$) on every other
site. 
Note however that in this case the values of the random kicks are refreshed 
at every plastic event\footnote{It has to be emphasized that when using a 
quenched kernel as in this case, there is a stability condition expressed in 
the fact that $G_{\bf q}$ has to be non-positive, otherwise we would obtain
exponentially growing modes. 
This is why we define the random kernel in ${\bf q}$ space. 
If we define a random kernel in real space instead, the negativity of ${\bf q}$ 
cannot be easily fulfilled.}.
From its very definition the mechanical noise in the HL model is a standard 
random walk, corresponding to a value of $\mu=2$.
In the quenched random case we are examining, we must first understand what are the 
properties of the uncorrelated mechanical noise felt by a particular target site.
The quenched random kernel $G_{\bf q}$ generates values $G_r$ that are mostly 
uncorrelated spatially, and distributed with a finite variance $\sigma$. 
This is enough to guarantee that we will find a value $\mu=2$ (and 
therefore~\cite{LinPRX2016} $\theta=1$) as in the HL model.
In addition, the dependence of $\sigma$ on the number of sites $N$ in the system is 
$\sigma\sim 1/\sqrt N$, as in the HL model.
Then we can write down the scaling of $P_0$ with $N$ from
(the limit of validity of) Eq.~\ref{us2h},
which is independent of the details of the kernel, as
\begin{equation}
    P_0\sim N^{-1/2}
\end{equation}
and also
\begin{equation}
    \langle \xmin\rangle \sim N^{-1/2}
\end{equation}
thus finding in the present case a different scaling 
that the one given by Eqs. \ref{eq:xmin_scaling} and \ref{eq:P0_scaling}.
The arguments that led to Eqs. \ref{eq:xmin_scaling} and \ref{eq:P0_scaling} 
fail here because the scaling of $\sigma$ with $N$ obtained in Eq. \ref{eq:sigma_xmin} was based in 
the conservation of the number of large 
kicks when system size is increased (see Fig. \ref{fig:sketch}), something that does not occur here because of the 
assumed non-decaying nature of the interactions.

We performed simulations with a quenched random kernel and evaluated
the distribution $P(x)$, and the value of $\langle \xmin\rangle$.
The simulations shown here were done in two spatial dimensions, but we verified
that exactly the same results are obtained in three dimensions if the number
of sites in the system is maintained.
This is of course related to the fact that in a randomly interacting model 
dimensionality plays no relevant role.

Figure \ref{fig:random} shows the results of simulations with the random kernel. 
The upper panel shows the form of $P(x)$ for different values of $N$ and we can 
see that the value $\theta=1$ for $P(x)\sim x^\theta$ in an intermediate range of
$x$ is well established as $N$ increases. 
The lower panel shows the scaling of  $P_0$ and $\langle \xmin\rangle$ with $N$, where we observe
clearly the expected $\sim N^{-1/2}$ dependence.


\begin{thebibliography}{46}%
\makeatletter
\providecommand \@ifxundefined [1]{%
 \@ifx{#1\undefined}
}%
\providecommand \@ifnum [1]{%
 \ifnum #1\expandafter \@firstoftwo
 \else \expandafter \@secondoftwo
 \fi
}%
\providecommand \@ifx [1]{%
 \ifx #1\expandafter \@firstoftwo
 \else \expandafter \@secondoftwo
 \fi
}%
\providecommand \natexlab [1]{#1}%
\providecommand \enquote  [1]{``#1''}%
\providecommand \bibnamefont  [1]{#1}%
\providecommand \bibfnamefont [1]{#1}%
\providecommand \citenamefont [1]{#1}%
\providecommand \href@noop [0]{\@secondoftwo}%
\providecommand \href [0]{\begingroup \@sanitize@url \@href}%
\providecommand \@href[1]{\@@startlink{#1}\@@href}%
\providecommand \@@href[1]{\endgroup#1\@@endlink}%
\providecommand \@sanitize@url [0]{\catcode `\\12\catcode `\$12\catcode
  `\&12\catcode `\#12\catcode `\^12\catcode `\_12\catcode `\%12\relax}%
\providecommand \@@startlink[1]{}%
\providecommand \@@endlink[0]{}%
\providecommand \url  [0]{\begingroup\@sanitize@url \@url }%
\providecommand \@url [1]{\endgroup\@href {#1}{\urlprefix }}%
\providecommand \urlprefix  [0]{URL }%
\providecommand \Eprint [0]{\href }%
\providecommand \doibase [0]{http://dx.doi.org/}%
\providecommand \selectlanguage [0]{\@gobble}%
\providecommand \bibinfo  [0]{\@secondoftwo}%
\providecommand \bibfield  [0]{\@secondoftwo}%
\providecommand \translation [1]{[#1]}%
\providecommand \BibitemOpen [0]{}%
\providecommand \bibitemStop [0]{}%
\providecommand \bibitemNoStop [0]{.\EOS\space}%
\providecommand \EOS [0]{\spacefactor3000\relax}%
\providecommand \BibitemShut  [1]{\csname bibitem#1\endcsname}%
\let\auto@bib@innerbib\@empty
\bibitem [{\citenamefont {Carlson}\ \emph {et~al.}(1994)\citenamefont
  {Carlson}, \citenamefont {Langer},\ and\ \citenamefont
  {Shaw}}]{carlson1994dynamics}%
  \BibitemOpen
  \bibfield  {author} {\bibinfo {author} {\bibfnamefont {J.~M.}\ \bibnamefont
  {Carlson}}, \bibinfo {author} {\bibfnamefont {J.~S.}\ \bibnamefont {Langer}},
  \ and\ \bibinfo {author} {\bibfnamefont {B.~E.}\ \bibnamefont {Shaw}},\
  }\href {https://doi.org/10.1103/RevModPhys.66.657} {\bibfield  {journal}
  {\bibinfo  {journal} {Reviews of Modern Physics}\ }\textbf {\bibinfo {volume}
  {66}},\ \bibinfo {pages} {657} (\bibinfo {year} {1994})}\BibitemShut
  {NoStop}%
\bibitem [{\citenamefont {Fisher}(1998)}]{FisherPR1998}%
  \BibitemOpen
  \bibfield  {author} {\bibinfo {author} {\bibfnamefont {D.~S.}\ \bibnamefont
  {Fisher}},\ }\href {https://doi.org/10.1016/S0370-1573(98)00008-8} {\bibfield
   {journal} {\bibinfo  {journal} {Phys. Rep.}\ }\textbf {\bibinfo {volume}
  {301}},\ \bibinfo {pages} {113} (\bibinfo {year} {1998})}\BibitemShut
  {NoStop}%
\bibitem [{\citenamefont {Ferré}\ \emph {et~al.}(2013)\citenamefont {Ferré},
  \citenamefont {Metaxas}, \citenamefont {Mougin}, \citenamefont {Jamet},
  \citenamefont {Gorchon},\ and\ \citenamefont {Jeudy}}]{FerreCRP2013}%
  \BibitemOpen
  \bibfield  {author} {\bibinfo {author} {\bibfnamefont {J.}~\bibnamefont
  {Ferré}}, \bibinfo {author} {\bibfnamefont {P.~J.}\ \bibnamefont {Metaxas}},
  \bibinfo {author} {\bibfnamefont {A.}~\bibnamefont {Mougin}}, \bibinfo
  {author} {\bibfnamefont {J.-P.}\ \bibnamefont {Jamet}}, \bibinfo {author}
  {\bibfnamefont {J.}~\bibnamefont {Gorchon}}, \ and\ \bibinfo {author}
  {\bibfnamefont {V.}~\bibnamefont {Jeudy}},\ }\href {\doibase
  https://doi.org/10.1016/j.crhy.2013.08.001} {\bibfield  {journal} {\bibinfo
  {journal} {Comptes Rendus Physique}\ }\textbf {\bibinfo {volume} {14}},\
  \bibinfo {pages} {651 } (\bibinfo {year} {2013})},\ \bibinfo {note}
  {disordered systems / Systèmes désordonnés}\BibitemShut {NoStop}%
\bibitem [{\citenamefont {de~Gennes}(1985)}]{DeGennesRMP1985}%
  \BibitemOpen
  \bibfield  {author} {\bibinfo {author} {\bibfnamefont {P.~G.}\ \bibnamefont
  {de~Gennes}},\ }\href {\doibase 10.1103/RevModPhys.57.827} {\bibfield
  {journal} {\bibinfo  {journal} {Rev. Mod. Phys.}\ }\textbf {\bibinfo {volume}
  {57}},\ \bibinfo {pages} {827} (\bibinfo {year} {1985})}\BibitemShut
  {NoStop}%
\bibitem [{\citenamefont {Nicolas}\ \emph {et~al.}(2018)\citenamefont
  {Nicolas}, \citenamefont {Ferrero}, \citenamefont {Martens},\ and\
  \citenamefont {Barrat}}]{NicolasRMP2018}%
  \BibitemOpen
  \bibfield  {author} {\bibinfo {author} {\bibfnamefont {A.}~\bibnamefont
  {Nicolas}}, \bibinfo {author} {\bibfnamefont {E.~E.}\ \bibnamefont
  {Ferrero}}, \bibinfo {author} {\bibfnamefont {K.}~\bibnamefont {Martens}}, \
  and\ \bibinfo {author} {\bibfnamefont {J.-L.}\ \bibnamefont {Barrat}},\
  }\href {\doibase 10.1103/RevModPhys.90.045006} {\bibfield  {journal}
  {\bibinfo  {journal} {Rev. Mod. Phys.}\ }\textbf {\bibinfo {volume} {90}},\
  \bibinfo {pages} {045006} (\bibinfo {year} {2018})}\BibitemShut {NoStop}%
\bibitem [{\citenamefont {Maloney}\ and\ \citenamefont
  {Lema\^{\i}tre}(2004)}]{MaloneyPRL2004}%
  \BibitemOpen
  \bibfield  {author} {\bibinfo {author} {\bibfnamefont {C.}~\bibnamefont
  {Maloney}}\ and\ \bibinfo {author} {\bibfnamefont {A.}~\bibnamefont
  {Lema\^{\i}tre}},\ }\href {https://doi.org/10.1103/PhysRevLett.93.016001}
  {\bibfield  {journal} {\bibinfo  {journal} {Physical Review Letters}\
  }\textbf {\bibinfo {volume} {93}},\ \bibinfo {pages} {016001} (\bibinfo
  {year} {2004})}\BibitemShut {NoStop}%
\bibitem [{\citenamefont {Lerner}\ and\ \citenamefont
  {Procaccia}(2009)}]{LernerPRE2009}%
  \BibitemOpen
  \bibfield  {author} {\bibinfo {author} {\bibfnamefont {E.}~\bibnamefont
  {Lerner}}\ and\ \bibinfo {author} {\bibfnamefont {I.}~\bibnamefont
  {Procaccia}},\ }\href {\doibase 10.1103/PhysRevE.79.066109} {\bibfield
  {journal} {\bibinfo  {journal} {Phys. Rev. E}\ }\textbf {\bibinfo {volume}
  {79}},\ \bibinfo {pages} {066109} (\bibinfo {year} {2009})}\BibitemShut
  {NoStop}%
\bibitem [{\citenamefont {Karmakar}\ \emph
  {et~al.}(2010{\natexlab{a}})\citenamefont {Karmakar}, \citenamefont
  {Lerner},\ and\ \citenamefont {Procaccia}}]{Karmakar2010}%
  \BibitemOpen
  \bibfield  {author} {\bibinfo {author} {\bibfnamefont {S.}~\bibnamefont
  {Karmakar}}, \bibinfo {author} {\bibfnamefont {E.}~\bibnamefont {Lerner}}, \
  and\ \bibinfo {author} {\bibfnamefont {I.}~\bibnamefont {Procaccia}},\ }\href
  {\doibase 10.1103/PhysRevE.82.055103} {\bibfield  {journal} {\bibinfo
  {journal} {Phys. Rev. E}\ }\textbf {\bibinfo {volume} {82}},\ \bibinfo
  {pages} {055103} (\bibinfo {year} {2010}{\natexlab{a}})}\BibitemShut
  {NoStop}%
\bibitem [{\citenamefont {Karmakar}\ \emph
  {et~al.}(2010{\natexlab{b}})\citenamefont {Karmakar}, \citenamefont {Lerner},
  \citenamefont {Procaccia},\ and\ \citenamefont {Zylberg}}]{KarmakarPRE2010b}%
  \BibitemOpen
  \bibfield  {author} {\bibinfo {author} {\bibfnamefont {S.}~\bibnamefont
  {Karmakar}}, \bibinfo {author} {\bibfnamefont {E.}~\bibnamefont {Lerner}},
  \bibinfo {author} {\bibfnamefont {I.}~\bibnamefont {Procaccia}}, \ and\
  \bibinfo {author} {\bibfnamefont {J.}~\bibnamefont {Zylberg}},\ }\href
  {\doibase 10.1103/PhysRevE.82.031301} {\bibfield  {journal} {\bibinfo
  {journal} {Phys. Rev. E}\ }\textbf {\bibinfo {volume} {82}},\ \bibinfo
  {pages} {031301} (\bibinfo {year} {2010}{\natexlab{b}})}\BibitemShut
  {NoStop}%
\bibitem [{\citenamefont {Lin}\ \emph {et~al.}(2014{\natexlab{a}})\citenamefont
  {Lin}, \citenamefont {Saade}, \citenamefont {Lerner}, \citenamefont {Rosso},\
  and\ \citenamefont {Wyart}}]{LinEPL2014}%
  \BibitemOpen
  \bibfield  {author} {\bibinfo {author} {\bibfnamefont {J.}~\bibnamefont
  {Lin}}, \bibinfo {author} {\bibfnamefont {A.}~\bibnamefont {Saade}}, \bibinfo
  {author} {\bibfnamefont {E.}~\bibnamefont {Lerner}}, \bibinfo {author}
  {\bibfnamefont {A.}~\bibnamefont {Rosso}}, \ and\ \bibinfo {author}
  {\bibfnamefont {M.}~\bibnamefont {Wyart}},\ }\href
  {https://doi.org/10.1209/0295-5075/105/26003} {\bibfield  {journal} {\bibinfo
   {journal} {Europhysics Letters (EPL)}\ }\textbf {\bibinfo {volume} {105}},\
  \bibinfo {pages} {26003} (\bibinfo {year} {2014}{\natexlab{a}})}\BibitemShut
  {NoStop}%
\bibitem [{\citenamefont {Lin}\ \emph {et~al.}(2014{\natexlab{b}})\citenamefont
  {Lin}, \citenamefont {Lerner}, \citenamefont {Rosso},\ and\ \citenamefont
  {Wyart}}]{LinPNAS2014}%
  \BibitemOpen
  \bibfield  {author} {\bibinfo {author} {\bibfnamefont {J.}~\bibnamefont
  {Lin}}, \bibinfo {author} {\bibfnamefont {E.}~\bibnamefont {Lerner}},
  \bibinfo {author} {\bibfnamefont {A.}~\bibnamefont {Rosso}}, \ and\ \bibinfo
  {author} {\bibfnamefont {M.}~\bibnamefont {Wyart}},\ }\href
  {https://doi.org/10.1073/pnas.1406391111} {\bibfield  {journal} {\bibinfo
  {journal} {Proceedings of the National Academy of Sciences}\ }\textbf
  {\bibinfo {volume} {111}},\ \bibinfo {pages} {14382} (\bibinfo {year}
  {2014}{\natexlab{b}})}\BibitemShut {NoStop}%
\bibitem [{\citenamefont {Budrikis}\ \emph {et~al.}(2017)\citenamefont
  {Budrikis}, \citenamefont {Castellanos}, \citenamefont {Sandfeld},
  \citenamefont {Zaiser},\ and\ \citenamefont
  {Zapperi}}]{budrikis2015universality}%
  \BibitemOpen
  \bibfield  {author} {\bibinfo {author} {\bibfnamefont {Z.}~\bibnamefont
  {Budrikis}}, \bibinfo {author} {\bibfnamefont {D.~F.}\ \bibnamefont
  {Castellanos}}, \bibinfo {author} {\bibfnamefont {S.}~\bibnamefont
  {Sandfeld}}, \bibinfo {author} {\bibfnamefont {M.}~\bibnamefont {Zaiser}}, \
  and\ \bibinfo {author} {\bibfnamefont {S.}~\bibnamefont {Zapperi}},\ }\href
  {https://www.nature.com/articles/ncomms15928} {\bibfield  {journal} {\bibinfo
   {journal} {Nat. Comm.}\ }\textbf {\bibinfo {volume} {8}},\ \bibinfo {pages}
  {15928} (\bibinfo {year} {2017})}\BibitemShut {NoStop}%
\bibitem [{\citenamefont {Tyukodi}\ \emph {et~al.}(2016)\citenamefont
  {Tyukodi}, \citenamefont {Patinet}, \citenamefont {Roux},\ and\ \citenamefont
  {Vandembroucq}}]{tyukodi2015depinning}%
  \BibitemOpen
  \bibfield  {author} {\bibinfo {author} {\bibfnamefont {B.}~\bibnamefont
  {Tyukodi}}, \bibinfo {author} {\bibfnamefont {S.}~\bibnamefont {Patinet}},
  \bibinfo {author} {\bibfnamefont {S.}~\bibnamefont {Roux}}, \ and\ \bibinfo
  {author} {\bibfnamefont {D.}~\bibnamefont {Vandembroucq}},\ }\href
  {https://doi.org/10.1103/PhysRevE.93.063005} {\bibfield  {journal} {\bibinfo
  {journal} {Phys. Rev. E}\ }\textbf {\bibinfo {volume} {93}},\ \bibinfo
  {pages} {063005} (\bibinfo {year} {2016})}\BibitemShut {NoStop}%
\bibitem [{\citenamefont {Liu}\ \emph {et~al.}(2016)\citenamefont {Liu},
  \citenamefont {Ferrero}, \citenamefont {Puosi}, \citenamefont {Barrat},\ and\
  \citenamefont {Martens}}]{liu2015driving}%
  \BibitemOpen
  \bibfield  {author} {\bibinfo {author} {\bibfnamefont {C.}~\bibnamefont
  {Liu}}, \bibinfo {author} {\bibfnamefont {E.~E.}\ \bibnamefont {Ferrero}},
  \bibinfo {author} {\bibfnamefont {F.}~\bibnamefont {Puosi}}, \bibinfo
  {author} {\bibfnamefont {J.-L.}\ \bibnamefont {Barrat}}, \ and\ \bibinfo
  {author} {\bibfnamefont {K.}~\bibnamefont {Martens}},\ }\href {\doibase
  10.1103/PhysRevLett.116.065501} {\bibfield  {journal} {\bibinfo  {journal}
  {Phys. Rev. Lett.}\ }\textbf {\bibinfo {volume} {116}},\ \bibinfo {pages}
  {065501} (\bibinfo {year} {2016})}\BibitemShut {NoStop}%
\bibitem [{\citenamefont {Ferrero}\ and\ \citenamefont
  {Jagla}(2019{\natexlab{a}})}]{FerreroSM2019}%
  \BibitemOpen
  \bibfield  {author} {\bibinfo {author} {\bibfnamefont {E.~E.}\ \bibnamefont
  {Ferrero}}\ and\ \bibinfo {author} {\bibfnamefont {E.~A.}\ \bibnamefont
  {Jagla}},\ }\href {\doibase 10.1039/C9SM01073D} {\bibfield  {journal}
  {\bibinfo  {journal} {Soft Matter}\ }\textbf {\bibinfo {volume} {15}},\
  \bibinfo {pages} {9041} (\bibinfo {year} {2019}{\natexlab{a}})}\BibitemShut
  {NoStop}%
\bibitem [{\citenamefont {Ferrero}\ and\ \citenamefont
  {Jagla}(2019{\natexlab{b}})}]{FerreroPRL2019}%
  \BibitemOpen
  \bibfield  {author} {\bibinfo {author} {\bibfnamefont {E.~E.}\ \bibnamefont
  {Ferrero}}\ and\ \bibinfo {author} {\bibfnamefont {E.~A.}\ \bibnamefont
  {Jagla}},\ }\href {\doibase 10.1103/PhysRevLett.123.218002} {\bibfield
  {journal} {\bibinfo  {journal} {Phys. Rev. Lett.}\ }\textbf {\bibinfo
  {volume} {123}},\ \bibinfo {pages} {218002} (\bibinfo {year}
  {2019}{\natexlab{b}})}\BibitemShut {NoStop}%
\bibitem [{\citenamefont {Fern\'andez~Aguirre}\ and\ \citenamefont
  {Jagla}(2018)}]{FernandezAguirrePRE2018}%
  \BibitemOpen
  \bibfield  {author} {\bibinfo {author} {\bibfnamefont {I.}~\bibnamefont
  {Fern\'andez~Aguirre}}\ and\ \bibinfo {author} {\bibfnamefont {E.~A.}\
  \bibnamefont {Jagla}},\ }\href {\doibase 10.1103/PhysRevE.98.013002}
  {\bibfield  {journal} {\bibinfo  {journal} {Phys. Rev. E}\ }\textbf {\bibinfo
  {volume} {98}},\ \bibinfo {pages} {013002} (\bibinfo {year}
  {2018})}\BibitemShut {NoStop}%
\bibitem [{\citenamefont {Jagla}(2018)}]{jagla2018prandtl}%
  \BibitemOpen
  \bibfield  {author} {\bibinfo {author} {\bibfnamefont {E.~A.}\ \bibnamefont
  {Jagla}},\ }\href {\doibase 10.1088/1742-5468/aa9db2} {\bibfield  {journal}
  {\bibinfo  {journal} {Journal of Statistical Mechanics: Theory and
  Experiment}\ }\textbf {\bibinfo {volume} {2018}},\ \bibinfo {pages} {013401}
  (\bibinfo {year} {2018})}\BibitemShut {NoStop}%
\bibitem [{\citenamefont {Salerno}\ and\ \citenamefont
  {Robbins}(2013)}]{Salerno2013}%
  \BibitemOpen
  \bibfield  {author} {\bibinfo {author} {\bibfnamefont {K.~M.}\ \bibnamefont
  {Salerno}}\ and\ \bibinfo {author} {\bibfnamefont {M.~O.}\ \bibnamefont
  {Robbins}},\ }\href {https://doi.org/10.1103/PhysRevE.88.062206} {\bibfield
  {journal} {\bibinfo  {journal} {Phys. Rev. E}\ }\textbf {\bibinfo {volume}
  {88}},\ \bibinfo {pages} {062206} (\bibinfo {year} {2013})}\BibitemShut
  {NoStop}%
\bibitem [{\citenamefont {Lin}\ \emph {et~al.}(2015)\citenamefont {Lin},
  \citenamefont {Gueudr{\'e}}, \citenamefont {Rosso},\ and\ \citenamefont
  {Wyart}}]{lin2015criticality}%
  \BibitemOpen
  \bibfield  {author} {\bibinfo {author} {\bibfnamefont {J.}~\bibnamefont
  {Lin}}, \bibinfo {author} {\bibfnamefont {T.}~\bibnamefont {Gueudr{\'e}}},
  \bibinfo {author} {\bibfnamefont {A.}~\bibnamefont {Rosso}}, \ and\ \bibinfo
  {author} {\bibfnamefont {M.}~\bibnamefont {Wyart}},\ }\href
  {https://doi.org/10.1103/PhysRevLett.115.168001} {\bibfield  {journal}
  {\bibinfo  {journal} {Physical review letters}\ }\textbf {\bibinfo {volume}
  {115}},\ \bibinfo {pages} {168001} (\bibinfo {year} {2015})}\BibitemShut
  {NoStop}%
\bibitem [{\citenamefont {Lin}\ and\ \citenamefont {Wyart}(2016)}]{LinPRX2016}%
  \BibitemOpen
  \bibfield  {author} {\bibinfo {author} {\bibfnamefont {J.}~\bibnamefont
  {Lin}}\ and\ \bibinfo {author} {\bibfnamefont {M.}~\bibnamefont {Wyart}},\
  }\href {https://doi.org/10.1103/PhysRevX.6.011005} {\bibfield  {journal}
  {\bibinfo  {journal} {Physical Review X}\ }\textbf {\bibinfo {volume} {6}},\
  \bibinfo {pages} {011005} (\bibinfo {year} {2016})}\BibitemShut {NoStop}%
\bibitem [{\citenamefont {Hentschel}\ \emph {et~al.}(2015)\citenamefont
  {Hentschel}, \citenamefont {Jaiswal}, \citenamefont {Procaccia},\ and\
  \citenamefont {Sastry}}]{HentschelPRE2015}%
  \BibitemOpen
  \bibfield  {author} {\bibinfo {author} {\bibfnamefont {H.~G.~E.}\
  \bibnamefont {Hentschel}}, \bibinfo {author} {\bibfnamefont {P.~K.}\
  \bibnamefont {Jaiswal}}, \bibinfo {author} {\bibfnamefont {I.}~\bibnamefont
  {Procaccia}}, \ and\ \bibinfo {author} {\bibfnamefont {S.}~\bibnamefont
  {Sastry}},\ }\href {\doibase 10.1103/PhysRevE.92.062302} {\bibfield
  {journal} {\bibinfo  {journal} {Phys. Rev. E}\ }\textbf {\bibinfo {volume}
  {92}},\ \bibinfo {pages} {062302} (\bibinfo {year} {2015})}\BibitemShut
  {NoStop}%
\bibitem [{\citenamefont {Ji}\ \emph {et~al.}(2019)\citenamefont {Ji},
  \citenamefont {Popovi\ifmmode~\acute{c}\else \'{c}\fi{}}, \citenamefont
  {de~Geus}, \citenamefont {Lerner},\ and\ \citenamefont
  {Wyart}}]{WenchengPRE2019}%
  \BibitemOpen
  \bibfield  {author} {\bibinfo {author} {\bibfnamefont {W.}~\bibnamefont
  {Ji}}, \bibinfo {author} {\bibfnamefont {M.}~\bibnamefont
  {Popovi\ifmmode~\acute{c}\else \'{c}\fi{}}}, \bibinfo {author} {\bibfnamefont
  {T.~W.~J.}\ \bibnamefont {de~Geus}}, \bibinfo {author} {\bibfnamefont
  {E.}~\bibnamefont {Lerner}}, \ and\ \bibinfo {author} {\bibfnamefont
  {M.}~\bibnamefont {Wyart}},\ }\href {\doibase 10.1103/PhysRevE.99.023003}
  {\bibfield  {journal} {\bibinfo  {journal} {Phys. Rev. E}\ }\textbf {\bibinfo
  {volume} {99}},\ \bibinfo {pages} {023003} (\bibinfo {year}
  {2019})}\BibitemShut {NoStop}%
\bibitem [{\citenamefont {Shang}\ \emph {et~al.}(2020)\citenamefont {Shang},
  \citenamefont {Guan},\ and\ \citenamefont {Barrat}}]{ShangPNAS2020}%
  \BibitemOpen
  \bibfield  {author} {\bibinfo {author} {\bibfnamefont {B.}~\bibnamefont
  {Shang}}, \bibinfo {author} {\bibfnamefont {P.}~\bibnamefont {Guan}}, \ and\
  \bibinfo {author} {\bibfnamefont {J.-L.}\ \bibnamefont {Barrat}},\ }\href
  {\doibase 10.1073/pnas.1915070117} {\bibfield  {journal} {\bibinfo  {journal}
  {Proceedings of the National Academy of Sciences}\ }\textbf {\bibinfo
  {volume} {117}},\ \bibinfo {pages} {86} (\bibinfo {year} {2020})}\BibitemShut
  {NoStop}%
\bibitem [{\citenamefont {Ruscher}\ and\ \citenamefont
  {Rottler}(2020)}]{ruscher2019residual}%
  \BibitemOpen
  \bibfield  {author} {\bibinfo {author} {\bibfnamefont {C.}~\bibnamefont
  {Ruscher}}\ and\ \bibinfo {author} {\bibfnamefont {J.}~\bibnamefont
  {Rottler}},\ }\href {\doibase 10.1039/D0SM01155J} {\bibfield  {journal}
  {\bibinfo  {journal} {Soft Matter}\ }\textbf {\bibinfo {volume} {16}},\
  \bibinfo {pages} {8940} (\bibinfo {year} {2020})}\BibitemShut {NoStop}%
\bibitem [{\citenamefont {Lin}\ and\ \citenamefont {Wyart}(2018)}]{LinPRE2018}%
  \BibitemOpen
  \bibfield  {author} {\bibinfo {author} {\bibfnamefont {J.}~\bibnamefont
  {Lin}}\ and\ \bibinfo {author} {\bibfnamefont {M.}~\bibnamefont {Wyart}},\
  }\href {\doibase 10.1103/PhysRevE.97.012603} {\bibfield  {journal} {\bibinfo
  {journal} {Phys. Rev. E}\ }\textbf {\bibinfo {volume} {97}},\ \bibinfo
  {pages} {012603} (\bibinfo {year} {2018})}\BibitemShut {NoStop}%
\bibitem [{Note1()}]{Note1}%
  \BibitemOpen
  \bibinfo {note} {As a matter of fact, \protect \emph {different} values for
  the exponent $\theta $ are presented in Ref.~\cite {LinPNAS2014} when either
  fitted from the $P(x)$ distribution or computed from the `extremal dynamics'
  (the scaling of $\left < x_\protect \text {min} \right >$) through Eq.\ref
  {eq:alphathetarelation}; ``a difference presumably resulting from corrections
  to scaling'' according to the authors.}\BibitemShut {Stop}%
\bibitem [{\citenamefont {Tyukodi}\ \emph {et~al.}(2019)\citenamefont
  {Tyukodi}, \citenamefont {Vandembroucq},\ and\ \citenamefont
  {Maloney}}]{TyukodiPRE2019}%
  \BibitemOpen
  \bibfield  {author} {\bibinfo {author} {\bibfnamefont {B.}~\bibnamefont
  {Tyukodi}}, \bibinfo {author} {\bibfnamefont {D.}~\bibnamefont
  {Vandembroucq}}, \ and\ \bibinfo {author} {\bibfnamefont {C.~E.}\
  \bibnamefont {Maloney}},\ }\href {\doibase 10.1103/PhysRevE.100.043003}
  {\bibfield  {journal} {\bibinfo  {journal} {Phys. Rev. E}\ }\textbf {\bibinfo
  {volume} {100}},\ \bibinfo {pages} {043003} (\bibinfo {year}
  {2019})}\BibitemShut {NoStop}%
\bibitem [{\citenamefont {Karimi}\ \emph {et~al.}(2017)\citenamefont {Karimi},
  \citenamefont {Ferrero},\ and\ \citenamefont {Barrat}}]{karimi2016inertia}%
  \BibitemOpen
  \bibfield  {author} {\bibinfo {author} {\bibfnamefont {K.}~\bibnamefont
  {Karimi}}, \bibinfo {author} {\bibfnamefont {E.~E.}\ \bibnamefont {Ferrero}},
  \ and\ \bibinfo {author} {\bibfnamefont {J.-L.}\ \bibnamefont {Barrat}},\
  }\href {\doibase 10.1103/PhysRevE.95.013003} {\bibfield  {journal} {\bibinfo
  {journal} {Phys. Rev. E}\ }\textbf {\bibinfo {volume} {95}},\ \bibinfo
  {pages} {013003} (\bibinfo {year} {2017})}\BibitemShut {NoStop}%
\bibitem [{\citenamefont {Lema{\^\i}tre}\ and\ \citenamefont
  {Caroli}(2007)}]{lemaitre_arxiv2007}%
  \BibitemOpen
  \bibfield  {author} {\bibinfo {author} {\bibfnamefont {A.}~\bibnamefont
  {Lema{\^\i}tre}}\ and\ \bibinfo {author} {\bibfnamefont {C.}~\bibnamefont
  {Caroli}},\ }\href {https://arxiv.org/abs/0705.3122} {\bibfield  {journal}
  {\bibinfo  {journal} {arXiv preprint arXiv:0705.3122}\ } (\bibinfo {year}
  {2007})}\BibitemShut {NoStop}%
\bibitem [{\citenamefont {Parley}\ \emph {et~al.}(2020)\citenamefont {Parley},
  \citenamefont {Fielding},\ and\ \citenamefont {Sollich}}]{parley2020aging}%
  \BibitemOpen
  \bibfield  {author} {\bibinfo {author} {\bibfnamefont {J.}~\bibnamefont
  {Parley}}, \bibinfo {author} {\bibfnamefont {S.}~\bibnamefont {Fielding}}, \
  and\ \bibinfo {author} {\bibfnamefont {P.}~\bibnamefont {Sollich}},\ }\href
  {https://arxiv.org/abs/2010.02593} {\bibfield  {journal} {\bibinfo  {journal}
  {arXiv preprint arXiv:2010.02593}\ } (\bibinfo {year} {2020})}\BibitemShut
  {NoStop}%
\bibitem [{Note2()}]{Note2}%
  \BibitemOpen
  \bibinfo {note} {In the simulations presented here the reinjection is made
  randomly and uniformly in the full interval $(0,1)$}\BibitemShut {NoStop}%
\bibitem [{Note3()}]{Note3}%
  \BibitemOpen
  \bibinfo {note} {We thank D. Vandembroucq for pointing this out.}\BibitemShut
  {Stop}%
\bibitem [{Note4()}]{Note4}%
  \BibitemOpen
  \bibinfo {note} {Note that instead of using at each step the latest
  ${x_{\protect \tt min}}$ to define the distribution $w(\xi )$, one could use
  the self-tuned mean value $\left <{x_{\protect \tt min}}\right >$ and the
  conclusions are identical.}\BibitemShut {Stop}%
\bibitem [{Note5()}]{Note5}%
  \BibitemOpen
  \bibinfo {note} {In fact the largest kicks are produced by neighbor
  avalanches, the coarse-grained lattice description imposes the upper cutoff
  of the kick distribution, the minimal distance.}\BibitemShut {Stop}%
\bibitem [{Note6()}]{Note6}%
  \BibitemOpen
  \bibinfo {note} {See Appendix B for the discussion of a case in which the
  assumptions made to derive this result do not apply, and then Eq.~\ref
  {eq:sigma_xmin} does not hold.}\BibitemShut {Stop}%
\bibitem [{\citenamefont {Picard}\ \emph {et~al.}(2005)\citenamefont {Picard},
  \citenamefont {Ajdari}, \citenamefont {Lequeux},\ and\ \citenamefont
  {Bocquet}}]{Picard2005}%
  \BibitemOpen
  \bibfield  {author} {\bibinfo {author} {\bibfnamefont {G.}~\bibnamefont
  {Picard}}, \bibinfo {author} {\bibfnamefont {A.}~\bibnamefont {Ajdari}},
  \bibinfo {author} {\bibfnamefont {F.}~\bibnamefont {Lequeux}}, \ and\
  \bibinfo {author} {\bibfnamefont {L.}~\bibnamefont {Bocquet}},\ }\href
  {https://doi.org/10.1103/PhysRevE.71.010501} {\bibfield  {journal} {\bibinfo
  {journal} {Physical Review E}\ }\textbf {\bibinfo {volume} {71}},\ \bibinfo
  {pages} {010501} (\bibinfo {year} {2005})}\BibitemShut {NoStop}%
\bibitem [{Note7()}]{Note7}%
  \BibitemOpen
  \bibinfo {note} {${x_{\protect \tt min}}$ is independently computed for each
  system size as the arithmetic average of the minimum $x$ values (in the
  $L\times L$ system) for each after-avalanche configuration in the steady
  state.}\BibitemShut {Stop}%
\bibitem [{\citenamefont {Jagla}(2020)}]{JaglaPRE2020}%
  \BibitemOpen
  \bibfield  {author} {\bibinfo {author} {\bibfnamefont {E.~A.}\ \bibnamefont
  {Jagla}},\ }\href {\doibase 10.1103/PhysRevE.101.043004} {\bibfield
  {journal} {\bibinfo  {journal} {Phys. Rev. E}\ }\textbf {\bibinfo {volume}
  {101}},\ \bibinfo {pages} {043004} (\bibinfo {year} {2020})}\BibitemShut
  {NoStop}%
\bibitem [{\citenamefont {Zhang}\ \emph {et~al.}(2020)\citenamefont {Zhang},
  \citenamefont {Ridout},\ and\ \citenamefont {Liu}}]{zhang2020interplay}%
  \BibitemOpen
  \bibfield  {author} {\bibinfo {author} {\bibfnamefont {G.}~\bibnamefont
  {Zhang}}, \bibinfo {author} {\bibfnamefont {S.}~\bibnamefont {Ridout}}, \
  and\ \bibinfo {author} {\bibfnamefont {A.~J.}\ \bibnamefont {Liu}},\ }\href
  {https://arxiv.org/abs/2009.11414} {\bibfield  {journal} {\bibinfo  {journal}
  {arXiv preprint arXiv:2009.11414}\ } (\bibinfo {year} {2020})}\BibitemShut
  {NoStop}%
\bibitem [{\citenamefont {Picard}\ \emph {et~al.}(2004)\citenamefont {Picard},
  \citenamefont {Ajdari}, \citenamefont {Lequeux},\ and\ \citenamefont
  {Bocquet}}]{Picard2004}%
  \BibitemOpen
  \bibfield  {author} {\bibinfo {author} {\bibfnamefont {G.}~\bibnamefont
  {Picard}}, \bibinfo {author} {\bibfnamefont {A.}~\bibnamefont {Ajdari}},
  \bibinfo {author} {\bibfnamefont {F.}~\bibnamefont {Lequeux}}, \ and\
  \bibinfo {author} {\bibfnamefont {L.}~\bibnamefont {Bocquet}},\ }\href
  {https://doi.org/10.1140/epje/i2004-10054-8} {\bibfield  {journal} {\bibinfo
  {journal} {The European physical journal. E, Soft matter}\ }\textbf {\bibinfo
  {volume} {15}},\ \bibinfo {pages} {371} (\bibinfo {year} {2004})}\BibitemShut
  {NoStop}%
\bibitem [{\citenamefont {Martens}\ \emph {et~al.}(2012)\citenamefont
  {Martens}, \citenamefont {Bocquet},\ and\ \citenamefont
  {Barrat}}]{Martens2012}%
  \BibitemOpen
  \bibfield  {author} {\bibinfo {author} {\bibfnamefont {K.}~\bibnamefont
  {Martens}}, \bibinfo {author} {\bibfnamefont {L.}~\bibnamefont {Bocquet}}, \
  and\ \bibinfo {author} {\bibfnamefont {J.-L.}\ \bibnamefont {Barrat}},\
  }\href {https://doi.org/10.1039/c2sm07090a} {\bibfield  {journal} {\bibinfo
  {journal} {Soft Matter}\ }\textbf {\bibinfo {volume} {8}},\ \bibinfo {pages}
  {4197} (\bibinfo {year} {2012})}\BibitemShut {NoStop}%
\bibitem [{\citenamefont {Nicolas}\ \emph {et~al.}(2014)\citenamefont
  {Nicolas}, \citenamefont {Martens},\ and\ \citenamefont
  {Barrat}}]{nicolas2014rheology}%
  \BibitemOpen
  \bibfield  {author} {\bibinfo {author} {\bibfnamefont {A.}~\bibnamefont
  {Nicolas}}, \bibinfo {author} {\bibfnamefont {K.}~\bibnamefont {Martens}}, \
  and\ \bibinfo {author} {\bibfnamefont {J.-L.}\ \bibnamefont {Barrat}},\
  }\href {https://doi.org/10.1209/0295-5075/107/44003} {\bibfield  {journal}
  {\bibinfo  {journal} {EPL (Europhysics Letters)}\ }\textbf {\bibinfo {volume}
  {107}},\ \bibinfo {pages} {44003} (\bibinfo {year} {2014})}\BibitemShut
  {NoStop}%
\bibitem [{\citenamefont {H\'{e}braud}\ and\ \citenamefont
  {Lequeux}(1998)}]{Hebraud1998}%
  \BibitemOpen
  \bibfield  {author} {\bibinfo {author} {\bibfnamefont {P.}~\bibnamefont
  {H\'{e}braud}}\ and\ \bibinfo {author} {\bibfnamefont {F.}~\bibnamefont
  {Lequeux}},\ }\href {https://doi.org/10.1103/PhysRevLett.81.2934} {\bibfield
  {journal} {\bibinfo  {journal} {Physical Review Letters}\ }\textbf {\bibinfo
  {volume} {81}},\ \bibinfo {pages} {2934} (\bibinfo {year}
  {1998})}\BibitemShut {NoStop}%
\bibitem [{\citenamefont {Agoritsas}\ \emph {et~al.}(2015)\citenamefont
  {Agoritsas}, \citenamefont {Bertin}, \citenamefont {Martens},\ and\
  \citenamefont {Barrat}}]{AgoritsasEPJE2015}%
  \BibitemOpen
  \bibfield  {author} {\bibinfo {author} {\bibfnamefont {E.}~\bibnamefont
  {Agoritsas}}, \bibinfo {author} {\bibfnamefont {E.}~\bibnamefont {Bertin}},
  \bibinfo {author} {\bibfnamefont {K.}~\bibnamefont {Martens}}, \ and\
  \bibinfo {author} {\bibfnamefont {J.-L.}\ \bibnamefont {Barrat}},\ }\href
  {\doibase 10.1140/epje/i2015-15071-x} {\bibfield  {journal} {\bibinfo
  {journal} {Eur. Phys. J. E}\ }\textbf {\bibinfo {volume} {38}},\ \bibinfo
  {pages} {71} (\bibinfo {year} {2015})}\BibitemShut {NoStop}%
\bibitem [{Note8()}]{Note8}%
  \BibitemOpen
  \bibinfo {note} {It has to be emphasized that when using a quenched kernel as
  in this case, there is a stability condition expressed in the fact that
  $G_{\protect \bf q}$ has to be non-positive, otherwise we would obtain
  exponentially growing modes. This is why we define the random kernel in
  ${\protect \bf q}$ space. If we define a random kernel in real space instead,
  the negativity of ${\protect \bf q}$ cannot be easily fulfilled.}\BibitemShut
  {Stop}%
\end{thebibliography}%

%

\end{document}